\documentclass[
 reprint,
 amsmath,amssymb,
 aps,
prb,
superscriptaddress]{revtex4-2}
\usepackage{array}
\usepackage{amsmath}
\usepackage{graphicx}
\usepackage{dcolumn}
\usepackage{bm}
\usepackage{float}
\usepackage{hyperref}
\hypersetup{colorlinks,linkcolor={blue},citecolor={blue},urlcolor={blue}}  
\usepackage{xtab,afterpage,longtable}
\makeatletter
\def\LT@LR@e{\LTleft\z@   \LTright\z@}
\makeatother
\newcommand{\hamiltonian}[1]{\mathcal{H}_\text{#1}}
\begin{document}

\preprint{APS/123-QED}

\title{Engineering non-Hermitian Second Order Topological Insulator in Quasicrystals}
\author{Chakradhar Rangi}
\email{crangi1@lsu.edu}
\affiliation{Department of Physics and Astronomy, Louisiana State University, Baton Rouge, LA 70803, USA}
\author{Ka-Ming Tam}
\affiliation{Department of Physics and Astronomy, Louisiana State University, Baton Rouge, LA 70803, USA}
\affiliation{Center for Computation and Technology, Louisiana State University, Baton Rouge, LA 70803, USA}
\author{Juana Moreno}
\affiliation{Department of Physics and Astronomy, Louisiana State University, Baton Rouge, LA 70803, USA}
\affiliation{Center for Computation and Technology, Louisiana State University, Baton Rouge, LA 70803, USA}
 
\date{\today}

\begin{abstract}

Non-Hermitian topological phases have gained immense attention due to their potential to unlock novel features beyond Hermitian bounds. PT-symmetric (Parity Time-reversal symmetric) non-Hermitian models have been studied extensively over the past decade. In recent years, the topological properties of general non-Hermitian models, regardless of the balance between gains and losses, have also attracted vast attention. Here we propose a non-Hermitian second-order topological (SOT) insulator that hosts gapless corner states on a two-dimensional quasi-crystalline lattice (QL). 
We first construct a non-Hermitian extension of the Bernevig-Hughes-Zhang (BHZ) model on a QL generated by the Amman-Beenker (AB) tiling. This model has real spectra and supports helical edge states. Corner states emerge  by adding a proper Wilson mass term that gaps out the edge states. We propose two variations of the mass term that result in fascinating characteristics. In the first variation, we obtain a purely real spectra for the second-order topological phase. In the latter, we get a complex spectra with corner states localized at only two corners. Our findings pave a path to engineering exotic SOT phases where corner states can be localized at designated corners.  
\end{abstract}
\maketitle

\section{\label{sec:intro}Introduction\protect}
\vspace{-0.25cm}
    Non-Hermitian topological phases are an exotic array of states which  represent a rapidly evolving field of study within condensed matter physics, optical science, and engineering~\cite{NHTP_Review_El-Ganainy2018, NHTP_Review_Gong2018, NHTP_Review_MartinezAlvarez2018, NHTP_Review_Ashida_2020, NHTP_Review_Ayan_2022, NHTP_Review_Bergholtz, NHTP_Review_Okuma2023}. While conventional Hermitian systems have long been the focus of research \cite{Hasan_Kane,Hasan_Moore,QiXiaoLiangReview,TIChiu,TIHaldane,TIKaneMele,TIBernevig,TIMarkus,TIMoore,FuKane,TIZhang}, the exploration of non-Hermitian phenomena has gained significant attention in recent years. This has been  motivated by viable theoretical and experimental platforms for realizing these exotic phases, such as Weyl semimetals \cite{WSM_Yong2017,WSM_Taiki2019,WSM_Kawabata2019,WSM_Haipin2022,WSM_LiKai2022,WSM_Tao2023}, models of finite quasiparticle lifetimes \cite{FLQP_Kozii2017,FLQP_Zyuzin2018Jan,FLQP_Shen2018Jul,FLQP_Yoshida2018Jul,FLQP_Papaj2019May,FLQP_McClarty2019Sep,FLQP_Hwan2022May,FLQP_Michen2022}, optical and mechanical systems subjected to gains and losses \cite{NHOptical_Makris2008, NHOptical_Chong2011,NHOptical_Regensburger2012,NHOptical_Hodaei2014, NHOptical_Peng2014,NHOptical_Jing2014, NHOptical_Liu2016,NHOptical_Lu2017,NHOptical_Soleymani2022,NHOptical_Zhang2023Jan,NHOptical_Arkhipov2023}, electrical circuits~\cite{EC_Helbig2020,EC_Chen2023,EC_Maopeng2023}, and even biological systems \cite{Bio_Nelson1998,Bio_Amir2016,Bio_Murugan2017}. 
    
    While the interplay of non-Hermiticity and topology has extended the understanding of their Hermitian counterparts, the non-Hermitian topological phases exhibit novel and richer features with no Hermitian counterparts. Some of the prominent examples include the existence of exceptional points (EPs) where more than one eigenstate coalesces \cite{NHTP_Review_Bergholtz,NHTP_Review_Okuma2023,EPs_Bender2007,EPs_Heiss2012}, and the bi-orthogonal bulk-boundary correspondence accompanied by non-Hermitian skin effects~\cite{BOBBC_Yao2018,BOBBC_Kunst2018,NHTP_Review_MartinezAlvarez2018, NHTP_Review_Okuma2023, Skin_Lee2016,Skin_Xiong2018,Skin_Yao2018}. These systems also extend the general symmetry classification of topological phases \cite{Bernard2002,HZhou_Table_nonHermitianSymmetries,KKawabata_prx_nonHermitianSymm,KKawabata_Unification_TRPT,KentaEsaki_Edge_states_TR,SymmClassification_Budich2019}.
    \begin{figure}[t!]
        \centering
        \begin{minipage}{0.5\columnwidth}
            \centering
            \includegraphics[width=\columnwidth]{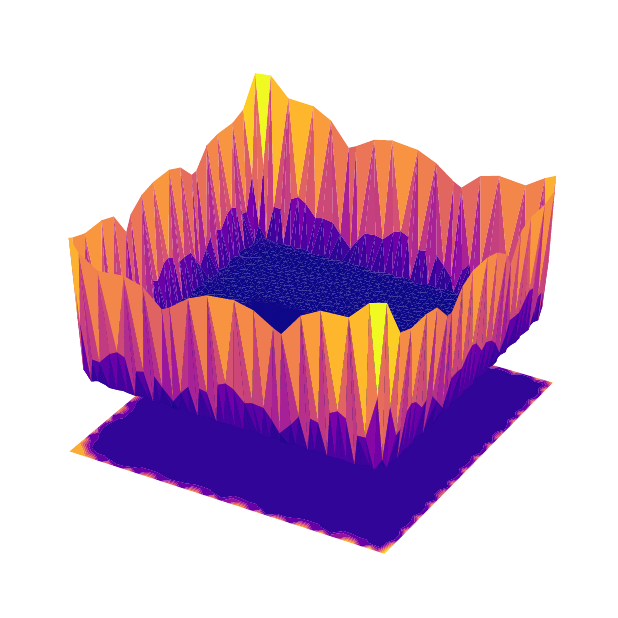} 
            \label{fig:3dPlotEdges}
        \end{minipage}\hfill
        \begin{minipage}{0.5\columnwidth}
            \centering
            \includegraphics[width=\columnwidth]{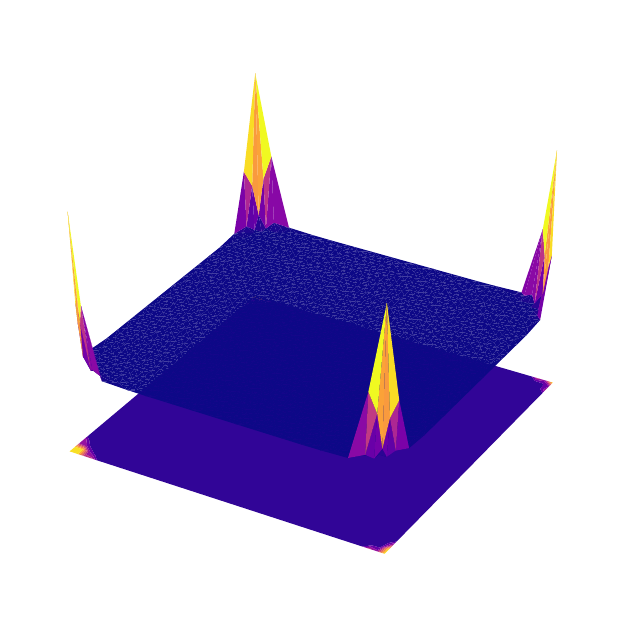} 
            \label{fig:3dPlotCorners}
        \end{minipage}\vspace{-2.5\baselineskip}
        \begin{minipage}{0.5\columnwidth}
            \centering
            \includegraphics[width=\columnwidth]{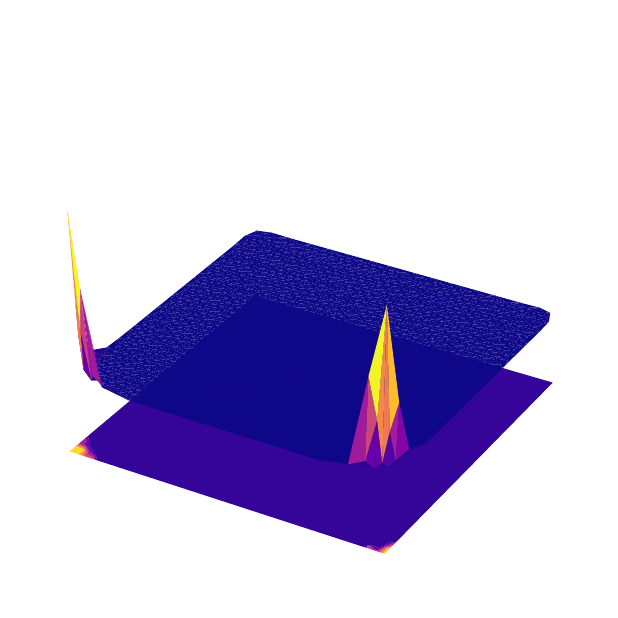} 
            \label{fig:3dtwocorners}
        \end{minipage}\hfill
        \begin{minipage}{0.5\columnwidth}
            \centering
            \includegraphics[width=\columnwidth]{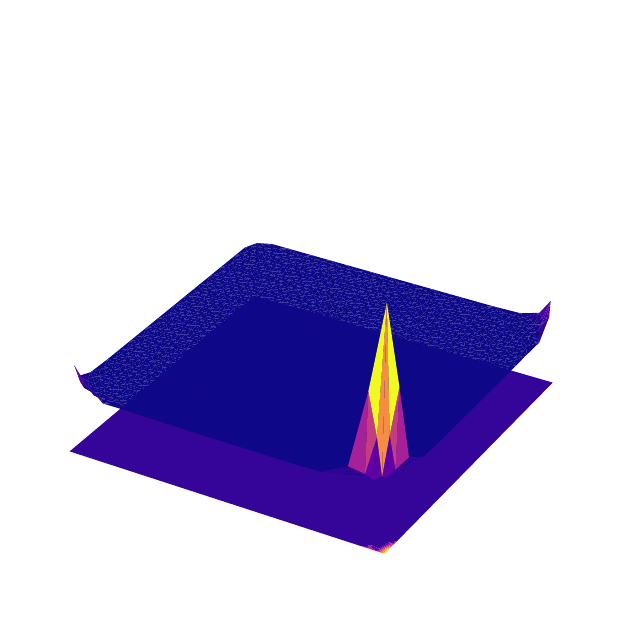} 
            \label{fig:3donecorner}
        \end{minipage}
        \vspace{-.75cm}
        \caption{Real-space wavefunction probability density of the topological states for several models. Top left: non-Hermitian extension of the BHZ model ($\hamiltonian{NH-BHZ}$),  Eq.~\eqref{eqn:nonhermitian_bhz_hamiltonian}.  Top right: non-Hermitian second order TI ($\hamiltonian{NH-SOTI}=\hamiltonian{NH-BHZ} + \hamiltonian{M}$), Eq.~\eqref{eq:NH-SOTI}. Bottom left: non-Hermitian BHZ with modified hopping term ($\hamiltonian{NH-BHZ} + \hamiltonian{M}''$), Eqs.~\eqref{eqn:nonhermitian_bhz_hamiltonian} and \eqref{eq:HM''}. Bottom right: non-Hermitian BHZ with a different modified hopping term ($\hamiltonian{NH-BHZ} + \hamiltonian{M}'''$), Eqs.~\eqref{eqn:nonhermitian_bhz_hamiltonian}
    and \eqref{eq:HM'''}. Parameters are $t_1=t_2=M=g=1.0$, and $\gamma=0.5$. The top left panel describes the wavefunction probability density of a typical in-gap state and the two-dimensional projection plot of its edge states for a non-Hermitian extension of the BHZ model defined on a quasicrsytalline lattice (Eq.~\eqref{eqn:nonhermitian_bhz_hamiltonian}).
    The top right panel describes the corner modes obtained by adding a mass term $\hamiltonian{M}$ to $\hamiltonian{NH-BHZ}$ (Eq.~\eqref{eq:NH-SOTI}) to gap out the edge states. The non-Hermiticity allows us to engineer more exotic SOT phases as shown in the two lower panels. }
        \label{fig:3dPlots}
    \end{figure}
    
Building upon the concept of topological insulators (TIs), the  notion of Hermitian higher-order topological insulators (HOTIs) has been proposed \cite{HOTI_Zhang2013,HOTI_Benalcazar2017,HOTI_Benalcazar2017Dec,HOTI_Josias2017,HOTI_Frank2018,HOTI_Xie2021}. Unlike conventional TIs, HOTIs have gapless states on lower-dimensional boundaries. For example, a second-order topological insulator (SOTI) in two dimensions hosts gapless corner modes, while a TI has gapless states on the whole boundary. Over the past few years, HOTIs have been discovered in aperiodic quasi-crystalline and amorphous systems \cite{Daniel-Varjas,RuiChen_etal_prl,AAgarwala_prr_HOTIAmorphou_}, expanding our understanding of topological phases in unconventional systems. 
    
Recently, Tao Liu et al. provided a framework to investigate non-Hermitian physics in HOTIs \cite{Liu-Tao}. They showed that 2D (3D) non-Hermitian crystalline insulators could host topologically protected second-order corner modes and, in contrast to their Hermitian counterpart, the gapless states can be localized only at one corner.  
    
    Motivated by these studies, we address whether it is possible to realize non-Hermitian HOTIs (NH-HOTIs) on quasicrystalline lattices (QLs). If these NH-HOTIs can be realized on QLs, is it possible to control and engineer them? In this work, we investigate non-Hermitian HOTIs on a 2D quasicrystalline square lattice generated by the Ammann-Beenker tiling pattern. We start with a non-Hermitian extension of the Bernevig-Hughes-Zhang (BHZ) model on a 2D quasicrystal respecting pseudo-hermiticity and reciprocity. We consider two variations of the Wilson-mass term to gap out the edge states resulting in corner states. Interestingly, we find that the NH-HOTI phase has purely real spectra in one case. The real spectrum of a non-Hermitian Hamiltonian is crucial in the context of dynamic stability. In contrast, we get complex spectra in the second case but observe unconventional phases where the corner modes can be localized at only one or two corners. This finding allows us to lay out a simple numerical approximation to understand and engineer the location of corner states. 
    
    The paper is organized as follows. In Sec. \ref{subsec:nonhermitianbhzmodel}, we define a non-Hermitian extension of the BHZ model that supports quantum spin Hall states (QSH) on a 2D quasicrystalline lattice. We consider two different mass terms that are added to this model. We analyze the spectrum and the resulting corner states of those models in Sec. \ref{subsec:SpectraAndCornerStates}. In Sec. \ref{subsec:TopoPhaseDiagram}, we compute the topological phase diagram and comment on the reality of the spectra. Sec. \ref{subsec:discussions} provides a summary and discussion. 
\section{Mass term induced corner modes in non-Hermitian BHZ model on QL}         \subsection{\label{subsec:nonhermitianbhzmodel}Model}
        Inspired by the non-Hermitian extension of the BHZ model on a square lattice~\cite{KKawabata_prr_realspectra}, we define a non-Hermitian BHZ (NH-BHZ) Hamiltonian on a 2D quasi-crystalline lattice. We consider a tight-binding non-Hermitian Hamiltonian on a 2D QL generated by the Ammann-Beenker (AB) tiling pattern, where the plane is tiled using squares and rhombi. Each lattice site consists of two orbitals. The second quantized Hamiltonian is given by
    \begin{align}\label{eqn:nonhermitian_bhz_hamiltonian}
            \hamiltonian{NH-BHZ} = \sum_{m\neq n} \hat{c}^\dagger_mH_{mn}\hat{c}_n + \sum_n \hat{c}^\dagger_nH_n \hat{c}_n,
        \end{align}
        where $\hat{c}^\dagger_n = (\hat{c}^\dagger_{n\alpha\uparrow},\hat{c}^\dagger_{n\alpha\downarrow},\hat{c}^\dagger_{n\beta\uparrow},\hat{c}^\dagger_{n\beta\downarrow})$ denotes the electron creation operator on site $n$; $\alpha$ and $\beta$ denote the orbital degrees of freedom at a given lattice site, and $\uparrow$ and $\downarrow$ represent the spin degrees of freedom. The hopping part of the Hamiltonian is
        \begin{equation}
        \begin{split}
        \label{non-Hermitian-hopping}
            H_{mn} &{}= - \frac{f(r_{mn})}{2}\Bigl[it_1\bigl(\sigma_3\tau_1\cos\phi_{mn} + \sigma_0\tau_2\sin\phi_{mn}\bigr) \\ &{}+ t_2 \sigma_0\tau_3 - \gamma \sigma_1\tau_1\cos\phi_{mn}\Bigr].
        \end{split}
        \end{equation}
        
\begin{figure}[htbp]
            \centering
            \includegraphics[width=0.6\columnwidth]{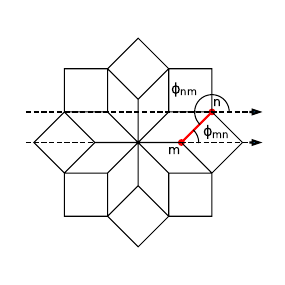}
            \vspace{-0.7cm}
            \caption{Illustration of the polar angle $\phi_{mn}$ defined on the QL as the angle made by the bond between sites $m$ and $n$ with respect to the horizontal. The bond between the sites $m$ and $n$ is marked red in color. The horizontal direction is depicted with dashed lines.
            }
            \label{fig:Angle}
    \end{figure}

        Here $t_1$ and $t_2$ are hopping amplitudes. The function $f(r_{mn}) \equiv \Theta(r_c - r_{mn})\text{exp}(1-r_{mn}/\xi)$ denotes the spatial decay factor of the hopping amplitude with $\xi$ as the decay length, and $r_{mn} = |\mathbf{r}_m-\mathbf{r}_n|$. The factor $\Theta(r_c - r_{mn})$ introduces a hard cut-off, $r_c$, for the hopping. $\sigma_i$ and $\tau_i (i=1,2,3)$ represent the Pauli matrices 
        acting on the spin and orbital sectors, respectively.   
        $\sigma_0$ and $\tau_0$ are the 2 $\times$ 2 identity matrices. 
        $\phi_{mn}$ represents the polar angle made by the bond between site $m$ and $n$ with respect to the horizontal direction as shown in Fig. \ref{fig:Angle} \cite{AAgarwala_TI_Amorphous,Mitchell2018_Amorphou_TI,AAgarwala_prr_HOTIAmorphou_}.
        As the factor $\cos(\phi_{mn})$ picks up a negative sign under $m\leftrightarrow n$, the last term in the above equation is the non-Hermitian part of the Hamiltonian. Consequently, $\gamma$ denotes the non-Hermitian strength. Physically, this results in an asymmetric hopping in our model. 
        
        The onsite term is given by
        \begin{align}\label{onsite_part_nonhermitianbhz}
            H_{n} = (M+2t_2)\sigma_0\tau_3,
        \end{align}
        where $M$ denotes the Dirac mass. Due to the distinction between conjugation and transposition in non-Hermitian Hamiltonians, the non-Hermiticity ramifies the internal symmetries extending the ten-fold Altland-Zirnbauer (AZ) symmetry of Hermitian systems to the 38-fold symmetry classes \cite{KKawabata_prx_nonHermitianSymm, HZhou_Table_nonHermitianSymmetries,Bernard2002}. The Hamiltonian in Eq.~\eqref{eqn:nonhermitian_bhz_hamiltonian} respects variants of time-reversal symmetry in non-Hermitian systems known as reciprocity \cite{KKawabata_prr_realspectra}, $\mathcal{T} \equiv i\sigma_2\tau_0$ and, pseudo-hermiticity, $\eta \equiv \sigma_3\tau_0$: 
        \begin{align}
            \mathcal{T}H^T\mathcal{T}^{-1} = H, \qquad \mathcal{T}\mathcal{T^*} = -1; \\ 
            \eta H^\dagger \eta^{-1} = H, \qquad \eta = \eta^{-1}.
        \end{align}
        where $H^T$ and $H^\dagger$ denote transposed and Hermitian conjugated Hamiltonian. A detailed symmetry analysis is carried out in Appendix A. 
        \begin{widetext}
        \begin{minipage}{\columnwidth}
            \begin{figure}[H]
            \centering
            \includegraphics[width=0.9\columnwidth]{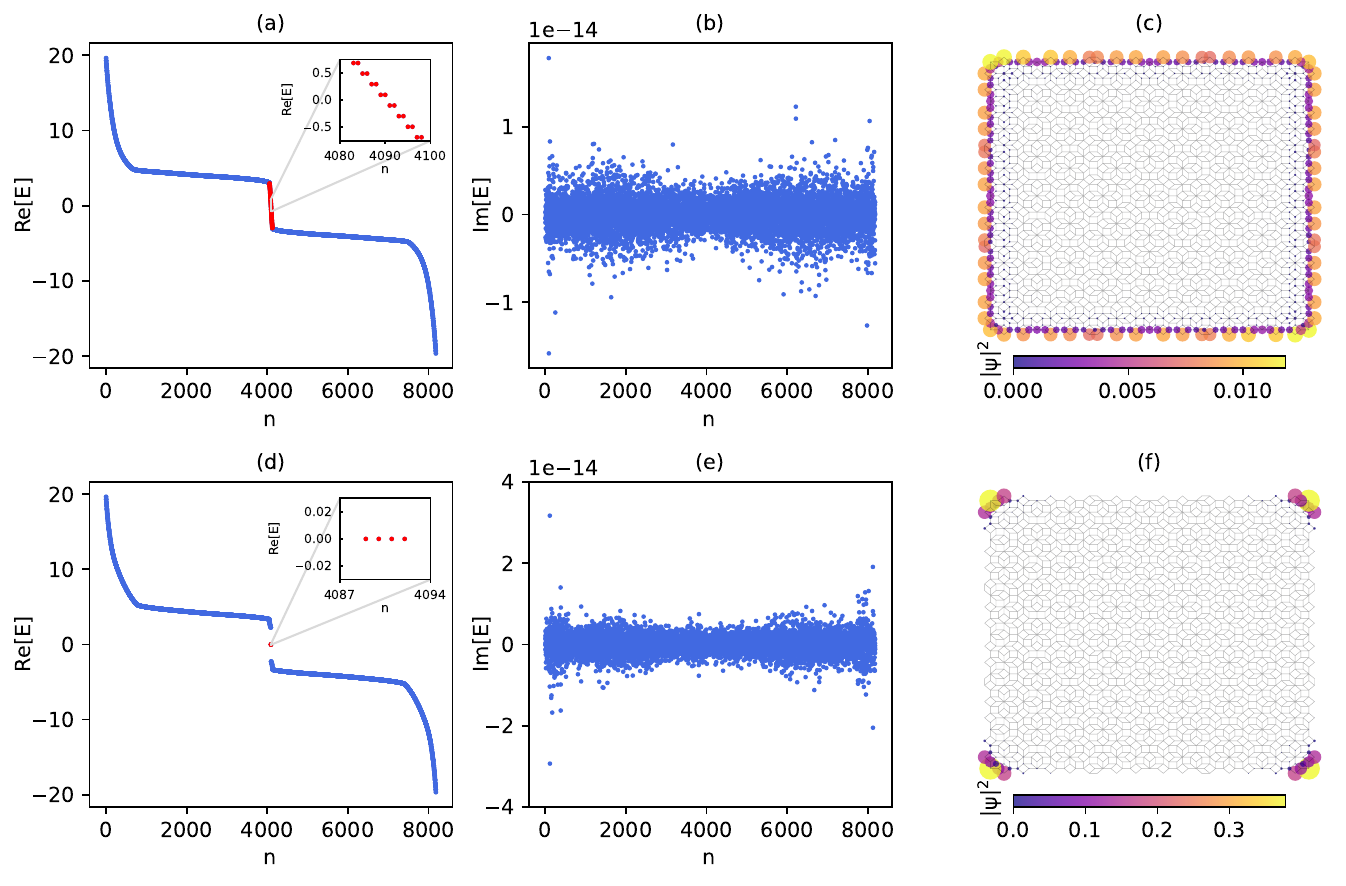}
            \vspace{-0.5cm}
            \caption{The complex energy spectrum and topological states of non-Hermitian-BHZ Hamiltonian (Eq.~\eqref{eqn:nonhermitian_bhz_hamiltonian}) and non-Hermitian SOTI Hamiltonian (Eq.~\eqref{eq:NH-SOTI})  with open boundary conditions. Panels (a) and (b) display the real and imaginary parts of the spectrum of $\hamiltonian{NH-BHZ}$
            (Eq.~\eqref{eqn:nonhermitian_bhz_hamiltonian})
            versus the eigenvalue index $n$  for $t_1 = t_2 = M = 1.0, \xi = 1.0$ and $\gamma = 0.5$.  The inset in panel (a) shows the in-gap QSH states of $\hamiltonian{NH-BHZ}$.
            Panel (c) displays the wavefunction probability density of a typical in-gap state distributed along the edges of the QL. 
            Panels (d) and (e) display the real and imaginary energies of $\hamiltonian{NH-SOTI}$ (Eq.~\eqref{eq:NH-SOTI}) 
            versus $n$ for for $t_1 = t_2 = M = 1.0, \xi = 1.0$, $\gamma = 0.5$ and $g = 1.0$. The inset in panel (d) show the zero-energy modes (ZEMs) of $\hamiltonian{NH-SOTI}$. Panel (f) displays the wavefunction probability density $\sum_{n\in \text{ZEMs}}|\psi_n|^2$ of the ZEMs localized on the corners of the QL. The number of sites in this simulation is 2,045.}
            \label{fig:SpectraNonHermitianBHZ}
            \end{figure}    
        \end{minipage}
        \end{widetext}
        \subsection{Spectrum and Corner States}\label{subsec:SpectraAndCornerStates}
        To obtain the quantum spin Hall (QSH) states and the spectrum, we diagonalize the $4N \times 4N$ Hamiltonian $\hamiltonian{NH-BHZ}$
            (Eq.~\eqref{eqn:nonhermitian_bhz_hamiltonian})
        defined on the QL under open boundary conditions (OBC) with $N$ denoting the number of sites and the following values for the parameters: $t_1 = t_2 = 1.0, M = 1.0, \xi = 1.0$ and $\gamma = 0.5$.  The spectrum and the probability distribution of the edge states are plotted in the top panels of Fig. \ref{fig:SpectraNonHermitianBHZ}. The bulk states are marked in blue as opposed to the in-gap states in red. A striking feature is that the spectrum is completely real, as evident from the imaginary part of the spectrum in Fig. \hyperref[fig:SpectraNonHermitianBHZ]{3(b)}. 
        The presence of pseudo-hermiticity symmetry, $\eta$, ensures that the bulk spectrum is real \cite{AliM_pseudo_antihermiticity,AliM_pseudo_antihermiticityII,AliM_pseudo_antihermiticityIII,KKawabata_prr_realspectra}. In addition, the combination of reciprocity and pseudo-hermiticity makes the edge states also to have a real spectra \cite{KKawabata_prr_realspectra}. In non-Hermitian systems, the reciprocity symmetry also leads to Kramer's degeneracy \cite{kramers_deg_Sato,KentaEsaki_Edge_states_TR,KKawabata_prx_nonHermitianSymm}. The inset of Fig. \ref{fig:SpectraNonHermitianBHZ}(a) shows a few in-gap states that are doubly degenerate as a consequence. These in-gap states live on the edges of the QL as indicated by the normalized probability density of a typical in-gap states displayed in Fig. \hyperref[fig:SpectraNonHermitianBHZ]{3(c)}.
        
        Now that we have designed a non-Hermitian QSH insulator on a QL, let us introduce a mass term in the Hamiltonian to gap out the in-gap states and obtain corner modes following the prescription given in Refs. \cite{RuiChen_etal_prl,AAgarwala_prr_HOTIAmorphou_}. We define:
        \begin{equation}\label{corner_mass_term}
            \hamiltonian{M} = \sum_{m \neq n} \hat{c}^\dagger_m \Bigr( \frac{f(r_{mn})}{2}g \sigma_2\tau_1 \cos(2\phi_{mn})\Bigl) \hat{c}_n,
        \end{equation}
        where $g$ is the magnitude of the Wilson mass and physically represents a hopping amplitude. Thus, the total Hamiltonian of a non-Hermitian second-order TI is 
        \begin{equation}\label{eq:NH-SOTI}
            \hamiltonian{NH-SOTI} = \hamiltonian{NH-BHZ} + \hamiltonian{M}.
        \end{equation}
        
        The mass term, $\hamiltonian{M}$ breaks the reciprocity symmetry and pseudo-hermiticity but preserves the chiral symmetry, 
        $\mathcal{S} = \mathcal{TC}$, with the non-Hermitian version defined as:
        \begin{equation}\label{chiral}
            \mathcal{S} H^\dagger \mathcal{S}^{-1} = -H.
        \end{equation}
    Additional details on the symmetry analysis are provided in Appendix A.
        
        We diagonalize the Hamiltonian $\hamiltonian{NH-SOTI}$ with  $g=1$ and the same set of parameters we used for $\hamiltonian{NH-BHZ}$
            (Eq.~\eqref{eqn:nonhermitian_bhz_hamiltonian}). The spectrum is plotted in panels \hyperref[fig:SpectraNonHermitianBHZ]{3(d)} and \hyperref[fig:SpectraNonHermitianBHZ]{3(e)}. In panel \hyperref[fig:SpectraNonHermitianBHZ]{3(d)}, we see that the in-gap states are gapped out, and four zero-energy modes (ZEMs) appear. An interesting feature is that the imaginary part of the spectrum is again zero. The corresponding ZEMs live on the corners of the QL, as evident from the probability distribution in panel \hyperref[fig:SpectraNonHermitianBHZ] {3(f)}.
        
\begin{figure}[tbp]
            \centering
            \begin{minipage}{0.5\columnwidth}
                \centering
                \includegraphics[width=\columnwidth]{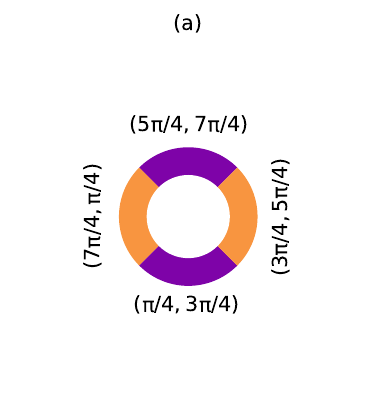} 
                \label{fig:SignMassTerm}
            \end{minipage}\hfill
            \begin{minipage}{0.5\columnwidth}
                \centering
                \includegraphics[width=\columnwidth]{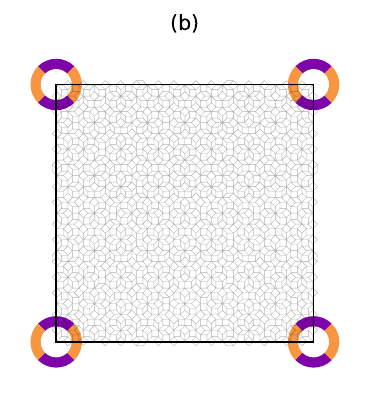}
                \label{fig:mechanismMassTerm}
            \end{minipage}
            \caption{Scheme using the generalized 
            Jackiw-Rebbi index theorem to account for the corner modes. Panel (a): a circular chart depicting the sign of the mass term, $\cos(2\theta_{edge})$, for $\theta_{edge} \in [0,2\pi)$. Orange and purple color denote positive and negative sign, respectively. The polar angle $\theta_{edge}$ made by the edge can be obtained by the angle at which the edge intersects the circular chart. Panel (b) illustrates the change in the sign of the mass term at each corner of the QL.  }
            \label{fig:mechanismMassChange}
        \end{figure}

        The appearance of these corner modes can be attributed to the generalized Jackiw-Rebbi (JR) index theorem \cite{Jackiw-Rebbi}, where the Wilson mass changes its sign. 
        To understand this, let us assume each edge of the QL to form a long bond \cite{RuiChen_etal_prl}. Since the mass term depends on the polar angle, $\phi_{mn}$, we can compare the angle each bond makes with the horizontal and obtain the sign of the term $\cos(2\phi_{mn})$. This is illustrated in Fig. \hyperref[fig:mechanismMassTerm] {4(b)} where the edges of the QL are approximated by a square.  The panel \hyperref[fig:mechanismMassChange]{4(a)} shows a circular chart, where the colors represent the sign of the mass term as a function of the polar angle of the edge, $\theta_{edge}$. 
        For example, the top right localized state in panel (b) will be formed by electrons flowing towards the right at the top horizontal edge and moving up along the right vertical edge. The right top horizontal edge forms an angle  $\theta_{edge}=0$, while the up right vertical edge forms an angle $\theta_{edge}=\pi/2$ with the horizontal axis.
        The label on each color section in Fig. \hyperref[fig:mechanismMassChange]{4(a)} represents the values of $\theta_{edge}$, at which the mass term changes sign. Effectively, the mass term, $\cos(2\theta_{edge})$, will distinguish the positive region, $\theta_{edge} \in [3\pi/4,5\pi/4] \bigcup \hspace{0.2em} [7\pi/4,\pi/4]$ and the negative region, $\theta_{edge} \in [\pi/4,3\pi/4] \bigcup \hspace{0.2em} [5\pi/4,7\pi/4]$. 
        At each corner of  Fig. \hyperref[fig:mechanismMassTerm] {4(b)}
        the adjacent sides of the square pass through orange (positive mass) and purple (negative mass) regions, indicating a mass domain wall resulting in a localized state. 

        \begin{widetext}
        \begin{minipage}{\columnwidth}
            \begin{figure}[H]
            \centering
            \includegraphics[width=0.62\columnwidth]{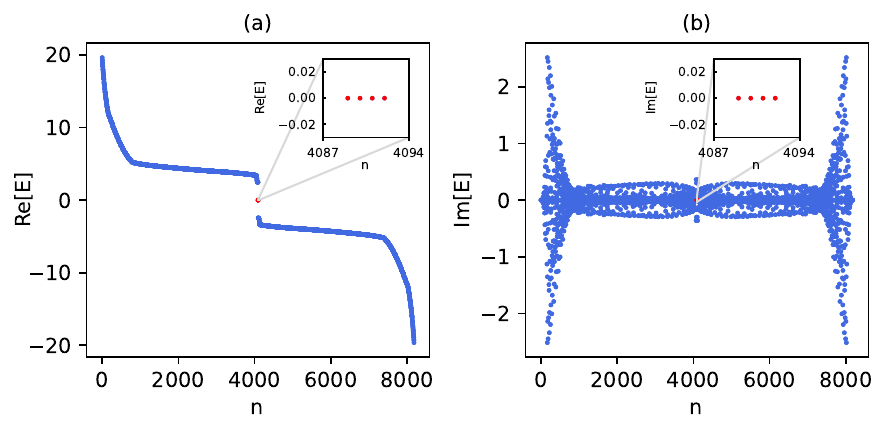}
            \includegraphics[width=0.27\columnwidth]{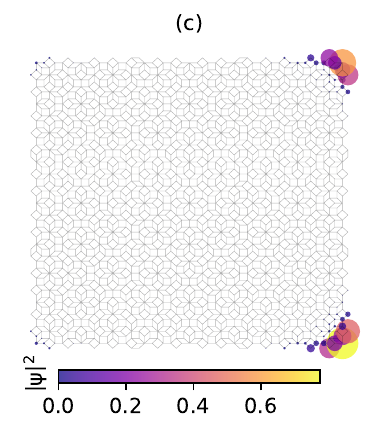} 
            \vspace{-0.5cm}
            \caption{The complex energy spectrum and corner states of          $\hamiltonian{NH-SOTI}'$ (Eq.~\ref{ruichen_mass_term})
            with $t_1 = t_2 = M = 1.0, \xi = 1.0$, $\gamma = 0.5$ and $g = 1.0$.
            Panels (a) and (b) display the real and imaginary parts of the spectrum, respectively. Insets show the ZEMs in red. Panel (c) shows the probability distribution of the ZEMs.  The number of sites is 2,045. }
            \label{fig:Ruichenmassterm}
            \end{figure}    
        \end{minipage}
        \end{widetext}
        
    We also study the mass term suggested in \cite{RuiChen_etal_prl}: 
        \begin{align}\label{ruichen_mass_term}
            \hamiltonian{M}' = \sum_{m \neq n} \hat{c}^\dagger_m \Bigr( \frac{f(r_{mn})}{2}g \sigma_1\tau_1 \cos(2\phi_{mn})\Bigl) \hat{c}_n, \nonumber \\
            \hamiltonian{NH-SOTI}' = \hamiltonian{NH-BHZ} + \hamiltonian{M}'.
        \end{align}
        The spectrum and the corner states of $\hamiltonian{NH-SOTI}'$ are plotted in Fig. \ref{fig:Ruichenmassterm} for $g=1.0, \gamma = 0.5$. 
        
        Comparing the spectrum and corner modes in Fig. \ref{fig:Ruichenmassterm} and \ref{fig:SpectraNonHermitianBHZ} we immediately notice the following differences: (i) the spectrum is complex in the former; (ii) the corner modes are localized at only two corners in Fig. \ref{fig:Ruichenmassterm}. We now shall address (ii) and resort to the next section to comment on (i). In short, we find that the interplay between the non-Hermitian asymmetric hopping term in Eq.~\eqref{non-Hermitian-hopping} and the mass term  in   Eq.~\ref{ruichen_mass_term},  $\hamiltonian{M}'$,
        plays a crucial role in dictating the number of corner modes. 
        
\begin{figure}[htbp]
            \centering
            \begin{minipage}{0.5\columnwidth}
                \centering
                \includegraphics[width=\columnwidth]{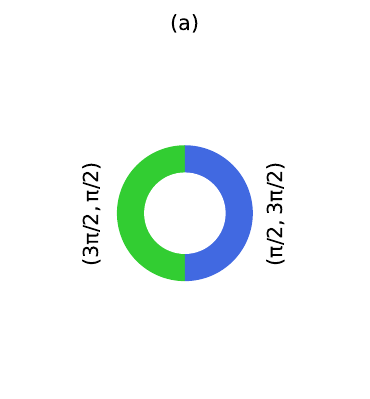} 
                \label{fig:DonutMassRatio}
            \end{minipage}\hfill
            \begin{minipage}{0.5\columnwidth}
                \centering
                \includegraphics[width=\columnwidth]{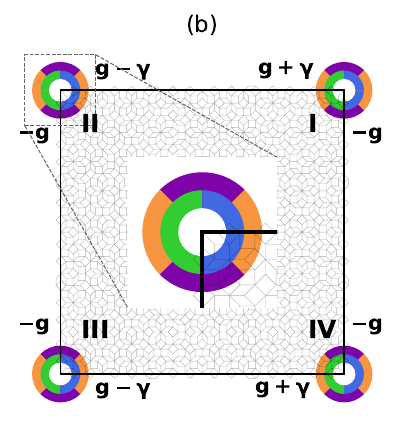} 
                \label{fig:HMMechanismCorners}
            \end{minipage}
            \vspace{-0.25cm}
            \caption{Mechanism describing the change in the ratio of the effective Wilson mass parameter, $\Tilde{g}=g\cos{2\theta_{edge}}+\gamma\cos{\theta_{edge}}$, at each corner. Panel (a): a circular chart displaying the sign of $\cos{\theta_{edge}}$ 
            as a function of the edge angle, $\theta_{edge}$, with blue and green sectors denoting negative and positive values, respectively. At the intersection of blue and green sectors, $\cos{\theta_{edge}}=0$ as $\theta_{edge} = \pi/2,3\pi/2$. Panel  (b): at each corner two circular charts are displayed, where the outer one is used to compute $g\cos{2\theta_{edge}}$, and the inner one is used to compute $\gamma\cos{\theta_{edge}}$. The term $\gamma\cos{\theta_{edge}}$ does not contribute to $\Tilde{g}$ for the vertical edges resulting in $\Tilde{g} = -g$. However, for the horizontal edges, $\theta_{edge}$ can take a value of either $0$ or $\pi$. At corners I and IV, $\theta_{edge}=0$ and thus, $\Tilde{g} = g+\gamma$ whereas at the other corners, $\theta_{edge} =\pi$ resulting in $\Tilde{g} = g-\gamma$. If the edge lies within the green sector, $\Tilde{g}$ increases, whereas $\Tilde{g}$ decreases if the edge lies within the blue sector.}
            \label{fig:MechanismMassRatio}
        \end{figure}

        The explanation for the apparent difference in the number of corner modes again employs the approximation scheme described in Fig. \ref{fig:mechanismMassChange}. 
        Since the mass term, $\hamiltonian{M}'$, and the non-Hermitian hopping term (Eq.~\eqref{non-Hermitian-hopping}) in $\hamiltonian{NH-BHZ}$  are both proportional to $\sigma_1\tau_1$, we expect both terms to contribute to the magnitude of the effective Wilson mass of each edge state.
         Note that we do not have an analytical expression for the effective Wilson mass due to the lack of translational symmetry, but we are able to give a crude estimate using the parameters of our model, namely, the Wilson mass parameter, $g$, and the non-Hermitian strength, $\gamma$. We assume that the strength of the effective Wilson mass at each edge roughly depends on $\Tilde{g} \equiv \bigl(g\cos{2\theta_{edge}} + \gamma\cos{\theta_{edge}}\bigr)$. For convenience, we call $\Tilde{g}$ the effective Wilson mass parameter. In Fig. \ref{fig:MechanismMassRatio}, we compute $\Tilde{g}$ at each corner with the help of the circular chart displayed in panel \hyperref[fig:DonutMassRatio]{6(a)}, which represents the behavior of $\cos{\theta_{edge}}$ as a function of $\theta_{edge}$. In particular, $\cos{\theta_{edge}}$ is positive when the edge intercepts the chart in the left half which is colored green. This corresponds to $\theta_{edge}$ in quadrants I and IV. On the other hand, $\cos{\theta_{edge}}$ is negative  when the edge intercepts the chart at the right ($\theta_{edge}$ in quadrants II and III). See Fig.~\hyperref[fig:DonutMassRatio]{6(a)}. 
         In Fig. \hyperref[fig:HMMechanismCorners]{6(b)}, we observe asymmetric values of $\Tilde{g}$ due to the non-Hermitian strength, $\gamma$. Namely, $\Tilde{g}$ for the horizontal edges at corners I and IV take a value of $g+\gamma$ whereas a value of $g-\gamma$ at corners II and III. As a result, the value of $\Tilde{g}$ at the horizontal edge over its value at the vertical edge increases at corners I and IV and decreases at corners II and III. Due to the increase in the ratio of the effective Wilson mass parameters at corners I and IV, the corresponding probability density of these  modes is enhanced, while the probability of a localized state at corners II and III is suppressed. This results in only two corners modes being observed. 
         The suppression of amplitude can be understood from the Jackiw-Rebbi solution of the Dirac equation. The JR solution for the wavefunction probability density with a mass domain at the origin depends on the masses as $m_1m_2/(m_1+m_2)$, where $m_1,m_2>0$. For a fixed mass $m_1$ the probability density only depends on the ratio $m_1/m_2$ as $1/(1+m_1/m_2)$ \cite{BookShen2017}. 
         
\begin{figure}[htbp]
            \centering
            \begin{minipage}{0.5\columnwidth}
                \centering
                \includegraphics[width=\columnwidth]{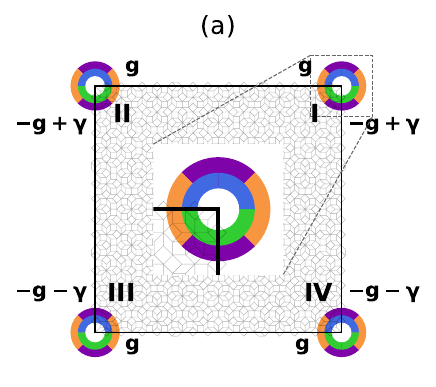} 
                \label{fig:H1Mechanism}
            \end{minipage}\hfill
            \begin{minipage}{0.5\columnwidth}
                \centering
                \includegraphics[width=\columnwidth]{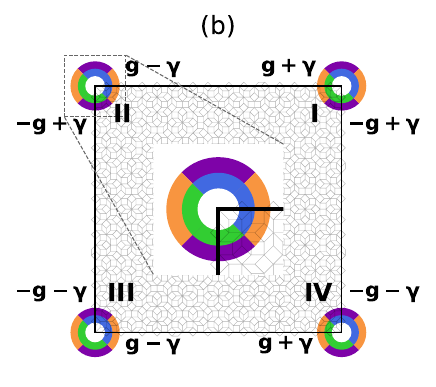} 
                \label{fig:H2Mechanism}
            \end{minipage}\vspace{-1.9\baselineskip}
            \vspace{-0.5cm}
            \begin{minipage}{0.5\columnwidth}
                \centering
                \includegraphics[width=\columnwidth]{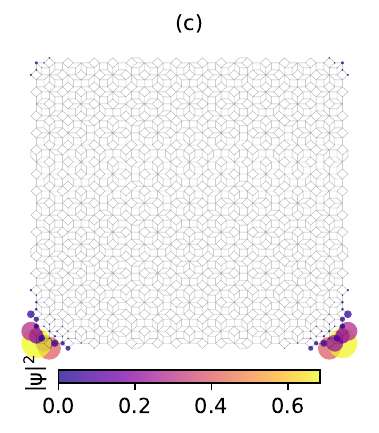} 
                \label{fig:H1corners}
            \end{minipage}\hfill
            \begin{minipage}{0.5\columnwidth}
                \centering
                \includegraphics[width=\columnwidth]{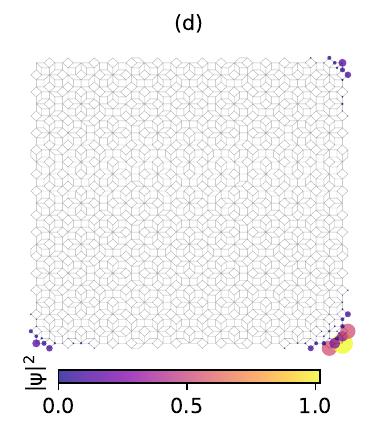} 
                \label{fig:H2corners}
            \end{minipage}
            \caption{Engineering corner states on QLs. (a) and (b) shows the change in $\Tilde{g}$ for the horizontal and vertical edges at all the corners. Note the change (c) and (d) shows the wavefunction probability density localized at two corners (III and IV) and a single corner (IV).}
            \label{fig:H1&H2Mechanism&Corners}
        \end{figure}

        This line of argument provides a remarkable approximation and guides our intuition in the numerical simulations. To demonstrate the utility of this approximation, we engineer a few scenarios for corner states by modifying the non-Hermitian hopping term of Eq. (\ref{non-Hermitian-hopping}) . We consider two different variations:
        \begin{gather}
            \hamiltonian{M}'' = \sum_{m \neq n} \hat{c}^\dagger_m H_1 \hat{c}_n \text{, where}  \\ H_1 = \frac{f(r_{mn})}{2}\gamma\sin{\phi_{mn}}\sigma_1\tau_1 \label{eq:HM''};\\ \hamiltonian{M}''' = \sum_{m \neq n} \hat{c}^\dagger_m H_2 \hat{c}_n \label{eq:HM'''}\text{, where} \\ H_2 = \frac{f(r_{mn})}{2}\gamma\bigl(\sin{\phi_{mn}}+\cos{\phi_{mn}}\bigr)\sigma_1\tau_1.
        \end{gather}     
         The term $H_1$ increases the ratio of $\Tilde{g}$ between the horizontal and vertical edges at corners III and IV and decreases it at corners I and II as described in Fig. \hyperref[fig:H1Mechanism]{7(a)}. 
         
        This results in localization of wavefunction probability density at corners III and IV as opposed to I and IV for $\hamiltonian{M}'$. $H_2$ produces an intriguing effect to produce a localized probability density at only corner IV. The corresponding wavefunction probability densities are shown in Figs. \hyperref[fig:H1corners]{7(c)} and \hyperref[fig:H2corners]{7(d)}, respectively.  A further inspection at Fig. \hyperref[fig:H2Mechanism]{7(b)} reveals that the ratio of $\Tilde{g}$ at corners II and IV are the same. This naturally leads to the question: Why do we see suppression of probability density at corner II as opposed to IV?
        We again invoke the JR solution for the wavefunction probability density, $m_1m_2/(m_1+m_2)$, which tells us that if both masses $m_1$ and $m_2$ decrease, the probability density is suppressed, explaining our observation at corner II compared to IV. Thus, this simple approximation scheme guides our intuition in engineering unique SOT phases. 
        
    \subsection{Topological Phase Diagram}\label{subsec:TopoPhaseDiagram}
     
\begin{figure}[t!]
            \centering
            \includegraphics[width=\columnwidth]{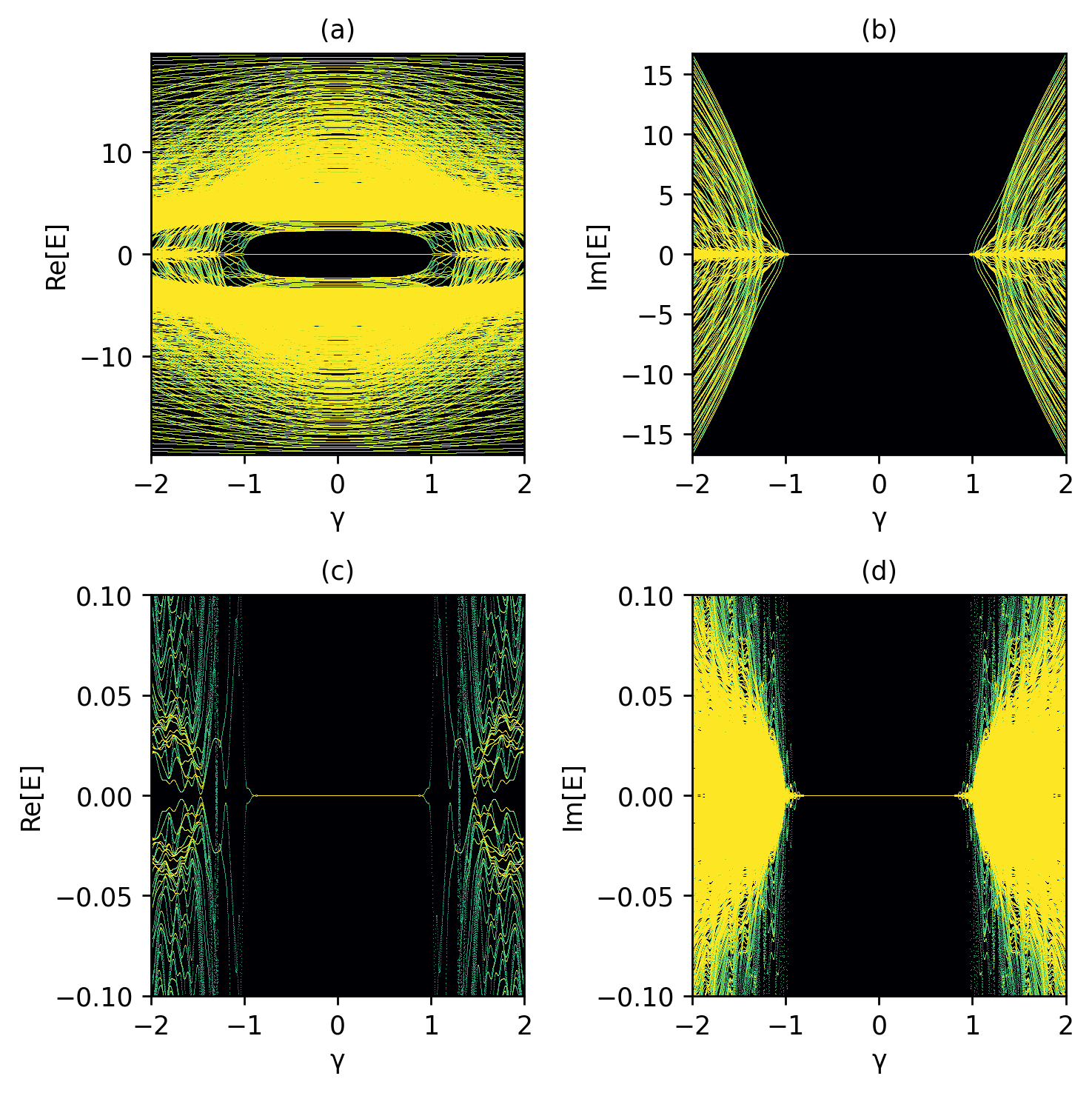}
            \caption{Topological phase transition for non-Hermitian model 
            $\hamiltonian{NH-SOTI}$ (Eq.~\ref{eq:NH-SOTI}) as a function of non-Hermitian strength, $\gamma$. Panels (a) and (b) show the evolution as function of $\gamma$ of the real and imaginary part of the spectra, respectively. Lighter color indicates larger density of states. Panels (c) and (d) show zoomed-in plots to track the evolution of the ZEMs.}
            \label{fig:phaseDiagramNonHBHZ}
    \end{figure}

    Figs. \hyperref[fig:SpectraNonHermitianBHZ]{3(d)} and \hyperref[fig:SpectraNonHermitianBHZ]{3(e)} revealed that the spectrum of $\hamiltonian{NH-SOTI}$ at $g=1.0$ and $\gamma=0.5$ is real. 
    We now ask if the reality of the spectrum is achieved only at one point or persists over a range of parameters.  To answer this question, we tweak the non-Hermitian parameter $\gamma$ over a range of values, $\gamma \in [-2,2]$, and plot the corresponding spectra as a function of $\gamma$. The values of other parameters in $\hamiltonian{NH-SOTI}$ remain the same. The results are plotted in Fig. \ref{fig:phaseDiagramNonHBHZ}. We witness a topological phase transition as we sweep $\gamma$ around $1.0$, where the ZEMs disappear and merge with the bulk bands. Another interesting characteristic of this transition is the disappearance of the real spectrum, as seen from the evolution of the imaginary part of the eigenenegies in Figs. \hyperref[fig:phaseDiagramNonHBHZ]{8(b)} and \hyperref[fig:phaseDiagramNonHBHZ]{8(d)}. 
    To understand the persistence of a real spectra over a finite range of $\gamma$, let us construct $\hamiltonian{NH-SOTI}$ on a 2D square lattice.
    This allow the use of analytical expressions for the spectrum in $k$-space, which we can then compare with $\hamiltonian{NH-SOTI}$ defined on a QL where there are not analytical expressions. The motivation for such an approach stems from the observation that in our numerical simulations for $\hamiltonian{NH-BHZ}$ on the QL, we recover the phase-diagram displayed in  Ref.~\cite{KKawabata_prr_realspectra} where $\hamiltonian{NH-BHZ}$ was defined on a square lattice. 
        The momentum space representation of $\hamiltonian{NH-SOTI}$ on a 2D square lattice is:
\begin{equation}\label{kSpaceHamiltonian}
            \begin{split}
                H_{\text{NH-SOTI}}(\mathbf{k}) &{}= t_1\bigl[ \sigma_3\tau_1 \sin{k_x} +\sigma_0\tau_2 \sin{k_y} \bigr] + \bigl[ M \\ &{}+ t_2(2 - \cos{k_x} - \cos{k_y}) \bigr]\sigma_0\tau_3  + g \times \\ &{}\bigr[\cos{k_x}- \cos{k_y} \bigl]\sigma_2\tau_1 + i\gamma\sin{k_x}\sigma_1\tau_1,  
            \end{split}
        \end{equation}
        and the corresponding eigenvalues, $E(\mathbf{k})$, can be computed as:
        \begin{equation}\label{energies_2D}
        \begin{split}
            E(\mathbf{k}) &{}= \pm \Bigl[t_1^2\sin^2{k_x} + t_1^2\sin^2{k_y} + \bigl(M + t_2[2 -\cos{k_x} \\ &{}- \cos{k_y}]\bigr)^2 + g^2\bigl(\cos{k_x} - \cos{k_y} \bigr)^2 - \gamma^2\sin{k_x}^2 \Bigr]^{\frac12}.
        \end{split}
        \end{equation}
        
        We recover the spectrum in Ref.~\cite{KKawabata_prr_realspectra} for $g=0$, $t_1 = -t_2$ up to a $k$-independent term in $t_2$. It is interesting to note that $E(\mathbf{k})$ in Eq.~\eqref{energies_2D} is either real or purely imaginary depending on the relative magnitudes of the parameters and $\mathbf{k}$. This is surprising as the addition of mass term breaks reciprocity and pseudo-Hermiticity which are the crucial symmetries responsible for the reality of the spectrum in the non-Hermitian BHZ model \cite{KKawabata_prr_realspectra}. 
        
        To compare we can construct the model of $\hamiltonian{NH-SOTI}'$ with $\hamiltonian{M}'$ as the mass term (Eq.~\ref{ruichen_mass_term})  and obtain the corresponding eigenvalues, $E'(\mathbf{k})$: 
        \begin{equation}\label{Hm'energies_2D}
        \begin{split}
            E'(\mathbf{k}) &{}= \pm \Bigl[t_1^2\sin^2{k_x} + t_1^2\sin^2{k_y} + \bigl(M + t_2[2 \\ &{}-\cos{k_x} - \cos{k_y}]\bigr)^2 + g^2\bigl(\cos{k_x} - \cos{k_y} \bigr)^2 \\ &{}- \gamma^2\sin{k_x}^2  + 2ig\gamma\sin{k_x}\bigl(\cos{k_x} - \cos{k_y} \bigr)\Bigr]^{\frac12}.
        \end{split}
        \end{equation}
        On comparing Eqs.~\eqref{Hm'energies_2D} and \eqref{energies_2D} we observe that the resulting spectrum is complex when the mass term is proportional to $\sigma_1\tau_1$.
        
\section{Discussion and Outlook}\label{subsec:discussions}
    We propose a non-Hermitian second-order topological phase on a 2D quasicrystalline lattice by adding two variations of a Wilson-mass term to a non-Hermitian extension of the BHZ model. In the former case, we find the spectrum to be purely real, which is important in the context of the dynamical stability of non-Hermitian systems. In the latter case, we find a complex spectrum, but the non-Hermiticity allow us to engineer more exotic SOT phases where localized states appear at only one or two corners. We also explore the reality of the spectra by comparing the eigenvalues of our models on the square lattice to those on the quasicrystalline lattice. To address whether such quasicrystals can be experimentally realized, one may consider avenues such as photonic quantum walks \cite{ExpQC_Weidemann2022,ExpQC_Quan2022}.
    
    Several open questions need to be addressed: (1) Is there a symmetry ensuring the reality of the spectra of the non-Hermitian second-order TI model we consider, $\hamiltonian{NH-SOTI}$ (Eq.~\ref{eq:NH-SOTI})? Even though the mass term $\hamiltonian{M}$ breaks pseudo-Hermiticity and reciprocity symmetry, which are crucial for the reality of spectra, we still end up with a real spectra. We do not find any obvious symmetry that is responsible for the real spectra. It would be interesting to further explore the reason behind this behavior. (2) Another question that arises in the context of topological phases is the nature of the topological invariant describing these SOT phases. We note that for the Hermitian case, it has been proposed that a topological invariant can be defined as a projection of the Hamiltonian from a higher dimension \cite{Daniel-Varjas, RuiChen_etal_prl}. It would be interesting to obtain the topological classification of non-Hermitian SOT in quasicrystals. Finally, it would be interesting to extend the study of non-Hermitian SOT phases to 3D quasicrystals.\\
    
    \begin{acknowledgments}
        We acknowledge Justin H. Wilson for helpful suggestions. This manuscript is based on work supported by the US Department of Energy, Office of Science, Office of Basic Energy Sciences, under Award Number DE-SC0017861. This work used high-performance computational resources provided by the Louisiana Optical Network Initiative and HPC@LSU computing.  
    \end{acknowledgments}
\bibliography{apssamp}

\providecommand{\noopsort}[1]{}\providecommand{\singleletter}[1]{#1}%
\begin{thebibliography}{83}%
\makeatletter
\providecommand \@ifxundefined [1]{%
 \@ifx{#1\undefined}
}%
\providecommand \@ifnum [1]{%
 \ifnum #1\expandafter \@firstoftwo
 \else \expandafter \@secondoftwo
 \fi
}%
\providecommand \@ifx [1]{%
 \ifx #1\expandafter \@firstoftwo
 \else \expandafter \@secondoftwo
 \fi
}%
\providecommand \natexlab [1]{#1}%
\providecommand \enquote  [1]{``#1''}%
\providecommand \bibnamefont  [1]{#1}%
\providecommand \bibfnamefont [1]{#1}%
\providecommand \citenamefont [1]{#1}%
\providecommand \href@noop [0]{\@secondoftwo}%
\providecommand \href [0]{\begingroup \@sanitize@url \@href}%
\providecommand \@href[1]{\@@startlink{#1}\@@href}%
\providecommand \@@href[1]{\endgroup#1\@@endlink}%
\providecommand \@sanitize@url [0]{\catcode `\\12\catcode `\$12\catcode
  `\&12\catcode `\#12\catcode `\^12\catcode `\_12\catcode `\%12\relax}%
\providecommand \@@startlink[1]{}%
\providecommand \@@endlink[0]{}%
\providecommand \url  [0]{\begingroup\@sanitize@url \@url }%
\providecommand \@url [1]{\endgroup\@href {#1}{\urlprefix }}%
\providecommand \urlprefix  [0]{URL }%
\providecommand \Eprint [0]{\href }%
\providecommand \doibase [0]{https://doi.org/}%
\providecommand \selectlanguage [0]{\@gobble}%
\providecommand \bibinfo  [0]{\@secondoftwo}%
\providecommand \bibfield  [0]{\@secondoftwo}%
\providecommand \translation [1]{[#1]}%
\providecommand \BibitemOpen [0]{}%
\providecommand \bibitemStop [0]{}%
\providecommand \bibitemNoStop [0]{.\EOS\space}%
\providecommand \EOS [0]{\spacefactor3000\relax}%
\providecommand \BibitemShut  [1]{\csname bibitem#1\endcsname}%
\let\auto@bib@innerbib\@empty
\bibitem [{\citenamefont {El-Ganainy}\ \emph {et~al.}(2018)\citenamefont
  {El-Ganainy}, \citenamefont {Makris}, \citenamefont {Khajavikhan},
  \citenamefont {Musslimani}, \citenamefont {Rotter},\ and\ \citenamefont
  {Christodoulides}}]{NHTP_Review_El-Ganainy2018}%
  \BibitemOpen
  \bibfield  {author} {\bibinfo {author} {\bibfnamefont {R.}~\bibnamefont
  {El-Ganainy}}, \bibinfo {author} {\bibfnamefont {K.~G.}\ \bibnamefont
  {Makris}}, \bibinfo {author} {\bibfnamefont {M.}~\bibnamefont {Khajavikhan}},
  \bibinfo {author} {\bibfnamefont {Z.~H.}\ \bibnamefont {Musslimani}},
  \bibinfo {author} {\bibfnamefont {S.}~\bibnamefont {Rotter}},\ and\ \bibinfo
  {author} {\bibfnamefont {D.~N.}\ \bibnamefont {Christodoulides}},\ }\href
  {https://doi.org/10.1038/nphys4323} {\bibfield  {journal} {\bibinfo
  {journal} {Nature Physics}\ }\textbf {\bibinfo {volume} {14}},\ \bibinfo
  {pages} {11} (\bibinfo {year} {2018})}\BibitemShut {NoStop}%
\bibitem [{\citenamefont {Gong}\ \emph {et~al.}(2018)\citenamefont {Gong},
  \citenamefont {Ashida}, \citenamefont {Kawabata}, \citenamefont {Takasan},
  \citenamefont {Higashikawa},\ and\ \citenamefont
  {Ueda}}]{NHTP_Review_Gong2018}%
  \BibitemOpen
  \bibfield  {author} {\bibinfo {author} {\bibfnamefont {Z.}~\bibnamefont
  {Gong}}, \bibinfo {author} {\bibfnamefont {Y.}~\bibnamefont {Ashida}},
  \bibinfo {author} {\bibfnamefont {K.}~\bibnamefont {Kawabata}}, \bibinfo
  {author} {\bibfnamefont {K.}~\bibnamefont {Takasan}}, \bibinfo {author}
  {\bibfnamefont {S.}~\bibnamefont {Higashikawa}},\ and\ \bibinfo {author}
  {\bibfnamefont {M.}~\bibnamefont {Ueda}},\ }\href
  {https://doi.org/10.1103/PhysRevX.8.031079} {\bibfield  {journal} {\bibinfo
  {journal} {Phys. Rev. X}\ }\textbf {\bibinfo {volume} {8}},\ \bibinfo {pages}
  {031079} (\bibinfo {year} {2018})}\BibitemShut {NoStop}%
\bibitem [{\citenamefont {Martinez~Alvarez}\ \emph {et~al.}(2018)\citenamefont
  {Martinez~Alvarez}, \citenamefont {Barrios~Vargas}, \citenamefont
  {Berdakin},\ and\ \citenamefont
  {Foa~Torres}}]{NHTP_Review_MartinezAlvarez2018}%
  \BibitemOpen
  \bibfield  {author} {\bibinfo {author} {\bibfnamefont {V.~M.}\ \bibnamefont
  {Martinez~Alvarez}}, \bibinfo {author} {\bibfnamefont {J.~E.}\ \bibnamefont
  {Barrios~Vargas}}, \bibinfo {author} {\bibfnamefont {M.}~\bibnamefont
  {Berdakin}},\ and\ \bibinfo {author} {\bibfnamefont {L.~E.~F.}\ \bibnamefont
  {Foa~Torres}},\ }\href {https://doi.org/10.1140/epjst/e2018-800091-5}
  {\bibfield  {journal} {\bibinfo  {journal} {The European Physical Journal
  Special Topics}\ }\textbf {\bibinfo {volume} {227}},\ \bibinfo {pages} {1295}
  (\bibinfo {year} {2018})}\BibitemShut {NoStop}%
\bibitem [{\citenamefont {Ashida}\ \emph {et~al.}(2020)\citenamefont {Ashida},
  \citenamefont {Gong},\ and\ \citenamefont {Ueda}}]{NHTP_Review_Ashida_2020}%
  \BibitemOpen
  \bibfield  {author} {\bibinfo {author} {\bibfnamefont {Y.}~\bibnamefont
  {Ashida}}, \bibinfo {author} {\bibfnamefont {Z.}~\bibnamefont {Gong}},\ and\
  \bibinfo {author} {\bibfnamefont {M.}~\bibnamefont {Ueda}},\ }\href
  {https://doi.org/10.1080/00018732.2021.1876991} {\bibfield  {journal}
  {\bibinfo  {journal} {Advances in Physics}\ }\textbf {\bibinfo {volume}
  {69}},\ \bibinfo {pages} {249} (\bibinfo {year} {2020})}\BibitemShut
  {NoStop}%
\bibitem [{\citenamefont {Banerjee}\ \emph {et~al.}(2023)\citenamefont
  {Banerjee}, \citenamefont {Sarkar}, \citenamefont {Dey},\ and\ \citenamefont
  {Narayan}}]{NHTP_Review_Ayan_2022}%
  \BibitemOpen
  \bibfield  {author} {\bibinfo {author} {\bibfnamefont {A.}~\bibnamefont
  {Banerjee}}, \bibinfo {author} {\bibfnamefont {R.}~\bibnamefont {Sarkar}},
  \bibinfo {author} {\bibfnamefont {S.}~\bibnamefont {Dey}},\ and\ \bibinfo
  {author} {\bibfnamefont {A.}~\bibnamefont {Narayan}},\ }\href
  {https://doi.org/10.1088/1361-648X/acd1cb} {\bibfield  {journal} {\bibinfo
  {journal} {Journal of Physics: Condensed Matter}\ }\textbf {\bibinfo {volume}
  {35}},\ \bibinfo {pages} {333001} (\bibinfo {year} {2023})}\BibitemShut
  {NoStop}%
\bibitem [{\citenamefont {Bergholtz}\ \emph {et~al.}(2021)\citenamefont
  {Bergholtz}, \citenamefont {Budich},\ and\ \citenamefont
  {Kunst}}]{NHTP_Review_Bergholtz}%
  \BibitemOpen
  \bibfield  {author} {\bibinfo {author} {\bibfnamefont {E.~J.}\ \bibnamefont
  {Bergholtz}}, \bibinfo {author} {\bibfnamefont {J.~C.}\ \bibnamefont
  {Budich}},\ and\ \bibinfo {author} {\bibfnamefont {F.~K.}\ \bibnamefont
  {Kunst}},\ }\href {https://doi.org/10.1103/RevModPhys.93.015005} {\bibfield
  {journal} {\bibinfo  {journal} {Rev. Mod. Phys.}\ }\textbf {\bibinfo {volume}
  {93}},\ \bibinfo {pages} {015005} (\bibinfo {year} {2021})}\BibitemShut
  {NoStop}%
\bibitem [{\citenamefont {Okuma}\ and\ \citenamefont
  {Sato}(2023)}]{NHTP_Review_Okuma2023}%
  \BibitemOpen
  \bibfield  {author} {\bibinfo {author} {\bibfnamefont {N.}~\bibnamefont
  {Okuma}}\ and\ \bibinfo {author} {\bibfnamefont {M.}~\bibnamefont {Sato}},\
  }\href {https://doi.org/10.1146/annurev-conmatphys-040521-033133} {\bibfield
  {journal} {\bibinfo  {journal} {Annual Review of Condensed Matter Physics}\
  }\textbf {\bibinfo {volume} {14}},\ \bibinfo {pages} {83} (\bibinfo {year}
  {2023})}\BibitemShut {NoStop}%
\bibitem [{\citenamefont {Hasan}\ and\ \citenamefont
  {Kane}(2010)}]{Hasan_Kane}%
  \BibitemOpen
  \bibfield  {author} {\bibinfo {author} {\bibfnamefont {M.~Z.}\ \bibnamefont
  {Hasan}}\ and\ \bibinfo {author} {\bibfnamefont {C.~L.}\ \bibnamefont
  {Kane}},\ }\href {https://doi.org/10.1103/RevModPhys.82.3045} {\bibfield
  {journal} {\bibinfo  {journal} {Rev. Mod. Phys.}\ }\textbf {\bibinfo {volume}
  {82}},\ \bibinfo {pages} {3045} (\bibinfo {year} {2010})}\BibitemShut
  {NoStop}%
\bibitem [{\citenamefont {Hasan}\ and\ \citenamefont
  {Moore}(2011)}]{Hasan_Moore}%
  \BibitemOpen
  \bibfield  {author} {\bibinfo {author} {\bibfnamefont {M.~Z.}\ \bibnamefont
  {Hasan}}\ and\ \bibinfo {author} {\bibfnamefont {J.~E.}\ \bibnamefont
  {Moore}},\ }\href {https://doi.org/10.1146/annurev-conmatphys-062910-140432}
  {\bibfield  {journal} {\bibinfo  {journal} {Annual Review of Condensed Matter
  Physics}\ }\textbf {\bibinfo {volume} {2}},\ \bibinfo {pages} {55} (\bibinfo
  {year} {2011})}\BibitemShut {NoStop}%
\bibitem [{\citenamefont {Qi}\ and\ \citenamefont
  {Zhang}(2011)}]{QiXiaoLiangReview}%
  \BibitemOpen
  \bibfield  {author} {\bibinfo {author} {\bibfnamefont {X.-L.}\ \bibnamefont
  {Qi}}\ and\ \bibinfo {author} {\bibfnamefont {S.-C.}\ \bibnamefont {Zhang}},\
  }\href {https://doi.org/10.1103/RevModPhys.83.1057} {\bibfield  {journal}
  {\bibinfo  {journal} {Rev. Mod. Phys.}\ }\textbf {\bibinfo {volume} {83}},\
  \bibinfo {pages} {1057} (\bibinfo {year} {2011})}\BibitemShut {NoStop}%
\bibitem [{\citenamefont {Chiu}\ \emph {et~al.}(2016)\citenamefont {Chiu},
  \citenamefont {Teo}, \citenamefont {Schnyder},\ and\ \citenamefont
  {Ryu}}]{TIChiu}%
  \BibitemOpen
  \bibfield  {author} {\bibinfo {author} {\bibfnamefont {C.-K.}\ \bibnamefont
  {Chiu}}, \bibinfo {author} {\bibfnamefont {J.~C.~Y.}\ \bibnamefont {Teo}},
  \bibinfo {author} {\bibfnamefont {A.~P.}\ \bibnamefont {Schnyder}},\ and\
  \bibinfo {author} {\bibfnamefont {S.}~\bibnamefont {Ryu}},\ }\href
  {https://doi.org/10.1103/RevModPhys.88.035005} {\bibfield  {journal}
  {\bibinfo  {journal} {Rev. Mod. Phys.}\ }\textbf {\bibinfo {volume} {88}},\
  \bibinfo {pages} {035005} (\bibinfo {year} {2016})}\BibitemShut {NoStop}%
\bibitem [{\citenamefont {Haldane}(1988)}]{TIHaldane}%
  \BibitemOpen
  \bibfield  {author} {\bibinfo {author} {\bibfnamefont {F.~D.~M.}\
  \bibnamefont {Haldane}},\ }\href
  {https://doi.org/10.1103/PhysRevLett.61.2015} {\bibfield  {journal} {\bibinfo
   {journal} {Phys. Rev. Lett.}\ }\textbf {\bibinfo {volume} {61}},\ \bibinfo
  {pages} {2015} (\bibinfo {year} {1988})}\BibitemShut {NoStop}%
\bibitem [{\citenamefont {Kane}\ and\ \citenamefont {Mele}(2005)}]{TIKaneMele}%
  \BibitemOpen
  \bibfield  {author} {\bibinfo {author} {\bibfnamefont {C.~L.}\ \bibnamefont
  {Kane}}\ and\ \bibinfo {author} {\bibfnamefont {E.~J.}\ \bibnamefont
  {Mele}},\ }\href {https://doi.org/10.1103/PhysRevLett.95.146802} {\bibfield
  {journal} {\bibinfo  {journal} {Phys. Rev. Lett.}\ }\textbf {\bibinfo
  {volume} {95}},\ \bibinfo {pages} {146802} (\bibinfo {year}
  {2005})}\BibitemShut {NoStop}%
\bibitem [{\citenamefont {Bernevig}\ \emph {et~al.}(2006)\citenamefont
  {Bernevig}, \citenamefont {Hughes},\ and\ \citenamefont
  {Zhang}}]{TIBernevig}%
  \BibitemOpen
  \bibfield  {author} {\bibinfo {author} {\bibfnamefont {B.~A.}\ \bibnamefont
  {Bernevig}}, \bibinfo {author} {\bibfnamefont {T.~L.}\ \bibnamefont
  {Hughes}},\ and\ \bibinfo {author} {\bibfnamefont {S.-C.}\ \bibnamefont
  {Zhang}},\ }\href {https://doi.org/10.1126/science.1133734} {\bibfield
  {journal} {\bibinfo  {journal} {Science}\ }\textbf {\bibinfo {volume}
  {314}},\ \bibinfo {pages} {1757} (\bibinfo {year} {2006})}\BibitemShut
  {NoStop}%
\bibitem [{\citenamefont {König}\ \emph {et~al.}(2007)\citenamefont {König},
  \citenamefont {Wiedmann}, \citenamefont {Brüne}, \citenamefont {Roth},
  \citenamefont {Buhmann}, \citenamefont {Molenkamp}, \citenamefont {Qi},\ and\
  \citenamefont {Zhang}}]{TIMarkus}%
  \BibitemOpen
  \bibfield  {author} {\bibinfo {author} {\bibfnamefont {M.}~\bibnamefont
  {König}}, \bibinfo {author} {\bibfnamefont {S.}~\bibnamefont {Wiedmann}},
  \bibinfo {author} {\bibfnamefont {C.}~\bibnamefont {Brüne}}, \bibinfo
  {author} {\bibfnamefont {A.}~\bibnamefont {Roth}}, \bibinfo {author}
  {\bibfnamefont {H.}~\bibnamefont {Buhmann}}, \bibinfo {author} {\bibfnamefont
  {L.~W.}\ \bibnamefont {Molenkamp}}, \bibinfo {author} {\bibfnamefont {X.-L.}\
  \bibnamefont {Qi}},\ and\ \bibinfo {author} {\bibfnamefont {S.-C.}\
  \bibnamefont {Zhang}},\ }\href {https://doi.org/10.1126/science.1148047}
  {\bibfield  {journal} {\bibinfo  {journal} {Science}\ }\textbf {\bibinfo
  {volume} {318}},\ \bibinfo {pages} {766} (\bibinfo {year}
  {2007})}\BibitemShut {NoStop}%
\bibitem [{\citenamefont {Moore}\ and\ \citenamefont
  {Balents}(2007)}]{TIMoore}%
  \BibitemOpen
  \bibfield  {author} {\bibinfo {author} {\bibfnamefont {J.~E.}\ \bibnamefont
  {Moore}}\ and\ \bibinfo {author} {\bibfnamefont {L.}~\bibnamefont
  {Balents}},\ }\href {https://doi.org/10.1103/PhysRevB.75.121306} {\bibfield
  {journal} {\bibinfo  {journal} {Phys. Rev. B}\ }\textbf {\bibinfo {volume}
  {75}},\ \bibinfo {pages} {121306} (\bibinfo {year} {2007})}\BibitemShut
  {NoStop}%
\bibitem [{\citenamefont {Fu}\ and\ \citenamefont {Kane}(2007)}]{FuKane}%
  \BibitemOpen
  \bibfield  {author} {\bibinfo {author} {\bibfnamefont {L.}~\bibnamefont
  {Fu}}\ and\ \bibinfo {author} {\bibfnamefont {C.~L.}\ \bibnamefont {Kane}},\
  }\href {https://doi.org/10.1103/PhysRevB.76.045302} {\bibfield  {journal}
  {\bibinfo  {journal} {Phys. Rev. B}\ }\textbf {\bibinfo {volume} {76}},\
  \bibinfo {pages} {045302} (\bibinfo {year} {2007})}\BibitemShut {NoStop}%
\bibitem [{\citenamefont {Zhang}\ \emph {et~al.}(2009)\citenamefont {Zhang},
  \citenamefont {Liu}, \citenamefont {Qi}, \citenamefont {Dai}, \citenamefont
  {Fang},\ and\ \citenamefont {Zhang}}]{TIZhang}%
  \BibitemOpen
  \bibfield  {author} {\bibinfo {author} {\bibfnamefont {H.}~\bibnamefont
  {Zhang}}, \bibinfo {author} {\bibfnamefont {C.-X.}\ \bibnamefont {Liu}},
  \bibinfo {author} {\bibfnamefont {X.-L.}\ \bibnamefont {Qi}}, \bibinfo
  {author} {\bibfnamefont {X.}~\bibnamefont {Dai}}, \bibinfo {author}
  {\bibfnamefont {Z.}~\bibnamefont {Fang}},\ and\ \bibinfo {author}
  {\bibfnamefont {S.-C.}\ \bibnamefont {Zhang}},\ }\href
  {https://doi.org/10.1038/nphys1270} {\bibfield  {journal} {\bibinfo
  {journal} {Nature Physics}\ }\textbf {\bibinfo {volume} {5}},\ \bibinfo
  {pages} {438} (\bibinfo {year} {2009})}\BibitemShut {NoStop}%
\bibitem [{\citenamefont {Xu}\ \emph {et~al.}(2017)\citenamefont {Xu},
  \citenamefont {Wang},\ and\ \citenamefont {Duan}}]{WSM_Yong2017}%
  \BibitemOpen
  \bibfield  {author} {\bibinfo {author} {\bibfnamefont {Y.}~\bibnamefont
  {Xu}}, \bibinfo {author} {\bibfnamefont {S.-T.}\ \bibnamefont {Wang}},\ and\
  \bibinfo {author} {\bibfnamefont {L.-M.}\ \bibnamefont {Duan}},\ }\href
  {https://doi.org/10.1103/PhysRevLett.118.045701} {\bibfield  {journal}
  {\bibinfo  {journal} {Phys. Rev. Lett.}\ }\textbf {\bibinfo {volume} {118}},\
  \bibinfo {pages} {045701} (\bibinfo {year} {2017})}\BibitemShut {NoStop}%
\bibitem [{\citenamefont {Matsushita}\ \emph {et~al.}(2019)\citenamefont
  {Matsushita}, \citenamefont {Nagai},\ and\ \citenamefont
  {Fujimoto}}]{WSM_Taiki2019}%
  \BibitemOpen
  \bibfield  {author} {\bibinfo {author} {\bibfnamefont {T.}~\bibnamefont
  {Matsushita}}, \bibinfo {author} {\bibfnamefont {Y.}~\bibnamefont {Nagai}},\
  and\ \bibinfo {author} {\bibfnamefont {S.}~\bibnamefont {Fujimoto}},\ }\href
  {https://doi.org/10.1103/PhysRevB.100.245205} {\bibfield  {journal} {\bibinfo
   {journal} {Phys. Rev. B}\ }\textbf {\bibinfo {volume} {100}},\ \bibinfo
  {pages} {245205} (\bibinfo {year} {2019})}\BibitemShut {NoStop}%
\bibitem [{\citenamefont {Kawabata}\ \emph
  {et~al.}(2019{\natexlab{a}})\citenamefont {Kawabata}, \citenamefont
  {Bessho},\ and\ \citenamefont {Sato}}]{WSM_Kawabata2019}%
  \BibitemOpen
  \bibfield  {author} {\bibinfo {author} {\bibfnamefont {K.}~\bibnamefont
  {Kawabata}}, \bibinfo {author} {\bibfnamefont {T.}~\bibnamefont {Bessho}},\
  and\ \bibinfo {author} {\bibfnamefont {M.}~\bibnamefont {Sato}},\ }\href
  {https://doi.org/10.1103/PhysRevLett.123.066405} {\bibfield  {journal}
  {\bibinfo  {journal} {Phys. Rev. Lett.}\ }\textbf {\bibinfo {volume} {123}},\
  \bibinfo {pages} {066405} (\bibinfo {year} {2019}{\natexlab{a}})}\BibitemShut
  {NoStop}%
\bibitem [{\citenamefont {Hu}\ \emph {et~al.}(2022)\citenamefont {Hu},
  \citenamefont {Zhao},\ and\ \citenamefont {Liu}}]{WSM_Haipin2022}%
  \BibitemOpen
  \bibfield  {author} {\bibinfo {author} {\bibfnamefont {H.}~\bibnamefont
  {Hu}}, \bibinfo {author} {\bibfnamefont {E.}~\bibnamefont {Zhao}},\ and\
  \bibinfo {author} {\bibfnamefont {W.~V.}\ \bibnamefont {Liu}},\ }\href
  {https://doi.org/10.1103/PhysRevB.106.094305} {\bibfield  {journal} {\bibinfo
   {journal} {Phys. Rev. B}\ }\textbf {\bibinfo {volume} {106}},\ \bibinfo
  {pages} {094305} (\bibinfo {year} {2022})}\BibitemShut {NoStop}%
\bibitem [{\citenamefont {Li}\ and\ \citenamefont {Xu}(2022)}]{WSM_LiKai2022}%
  \BibitemOpen
  \bibfield  {author} {\bibinfo {author} {\bibfnamefont {K.}~\bibnamefont
  {Li}}\ and\ \bibinfo {author} {\bibfnamefont {Y.}~\bibnamefont {Xu}},\ }\href
  {https://doi.org/10.1103/PhysRevLett.129.093001} {\bibfield  {journal}
  {\bibinfo  {journal} {Phys. Rev. Lett.}\ }\textbf {\bibinfo {volume} {129}},\
  \bibinfo {pages} {093001} (\bibinfo {year} {2022})}\BibitemShut {NoStop}%
\bibitem [{\citenamefont {Tao}\ \emph {et~al.}(2023)\citenamefont {Tao},
  \citenamefont {Qin},\ and\ \citenamefont {Xu}}]{WSM_Tao2023}%
  \BibitemOpen
  \bibfield  {author} {\bibinfo {author} {\bibfnamefont {Y.-L.}\ \bibnamefont
  {Tao}}, \bibinfo {author} {\bibfnamefont {T.}~\bibnamefont {Qin}},\ and\
  \bibinfo {author} {\bibfnamefont {Y.}~\bibnamefont {Xu}},\ }\href
  {https://doi.org/10.1103/PhysRevB.107.035140} {\bibfield  {journal} {\bibinfo
   {journal} {Phys. Rev. B}\ }\textbf {\bibinfo {volume} {107}},\ \bibinfo
  {pages} {035140} (\bibinfo {year} {2023})}\BibitemShut {NoStop}%
\bibitem [{\citenamefont {Kozii}\ and\ \citenamefont
  {Fu}(2017)}]{FLQP_Kozii2017}%
  \BibitemOpen
  \bibfield  {author} {\bibinfo {author} {\bibfnamefont {V.}~\bibnamefont
  {Kozii}}\ and\ \bibinfo {author} {\bibfnamefont {L.}~\bibnamefont {Fu}},\
  }\href@noop {} {} (\bibinfo {year} {2017}),\ \Eprint
  {https://arxiv.org/abs/1708.05841} {arXiv:1708.05841} \BibitemShut {NoStop}%
\bibitem [{\citenamefont {Zyuzin}\ and\ \citenamefont
  {Zyuzin}(2018)}]{FLQP_Zyuzin2018Jan}%
  \BibitemOpen
  \bibfield  {author} {\bibinfo {author} {\bibfnamefont {A.~A.}\ \bibnamefont
  {Zyuzin}}\ and\ \bibinfo {author} {\bibfnamefont {A.~Y.}\ \bibnamefont
  {Zyuzin}},\ }\href {https://doi.org/10.1103/PhysRevB.97.041203} {\bibfield
  {journal} {\bibinfo  {journal} {Phys. Rev. B}\ }\textbf {\bibinfo {volume}
  {97}},\ \bibinfo {pages} {041203} (\bibinfo {year} {2018})}\BibitemShut
  {NoStop}%
\bibitem [{\citenamefont {Shen}\ and\ \citenamefont
  {Fu}(2018)}]{FLQP_Shen2018Jul}%
  \BibitemOpen
  \bibfield  {author} {\bibinfo {author} {\bibfnamefont {H.}~\bibnamefont
  {Shen}}\ and\ \bibinfo {author} {\bibfnamefont {L.}~\bibnamefont {Fu}},\
  }\href {https://doi.org/10.1103/PhysRevLett.121.026403} {\bibfield  {journal}
  {\bibinfo  {journal} {Phys. Rev. Lett.}\ }\textbf {\bibinfo {volume} {121}},\
  \bibinfo {pages} {026403} (\bibinfo {year} {2018})}\BibitemShut {NoStop}%
\bibitem [{\citenamefont {Yoshida}\ \emph {et~al.}(2018)\citenamefont
  {Yoshida}, \citenamefont {Peters},\ and\ \citenamefont
  {Kawakami}}]{FLQP_Yoshida2018Jul}%
  \BibitemOpen
  \bibfield  {author} {\bibinfo {author} {\bibfnamefont {T.}~\bibnamefont
  {Yoshida}}, \bibinfo {author} {\bibfnamefont {R.}~\bibnamefont {Peters}},\
  and\ \bibinfo {author} {\bibfnamefont {N.}~\bibnamefont {Kawakami}},\ }\href
  {https://doi.org/10.1103/PhysRevB.98.035141} {\bibfield  {journal} {\bibinfo
  {journal} {Phys. Rev. B}\ }\textbf {\bibinfo {volume} {98}},\ \bibinfo
  {pages} {035141} (\bibinfo {year} {2018})}\BibitemShut {NoStop}%
\bibitem [{\citenamefont {Papaj}\ \emph {et~al.}(2019)\citenamefont {Papaj},
  \citenamefont {Isobe},\ and\ \citenamefont {Fu}}]{FLQP_Papaj2019May}%
  \BibitemOpen
  \bibfield  {author} {\bibinfo {author} {\bibfnamefont {M.}~\bibnamefont
  {Papaj}}, \bibinfo {author} {\bibfnamefont {H.}~\bibnamefont {Isobe}},\ and\
  \bibinfo {author} {\bibfnamefont {L.}~\bibnamefont {Fu}},\ }\href
  {https://doi.org/10.1103/PhysRevB.99.201107} {\bibfield  {journal} {\bibinfo
  {journal} {Phys. Rev. B}\ }\textbf {\bibinfo {volume} {99}},\ \bibinfo
  {pages} {201107} (\bibinfo {year} {2019})}\BibitemShut {NoStop}%
\bibitem [{\citenamefont {McClarty}\ and\ \citenamefont
  {Rau}(2019)}]{FLQP_McClarty2019Sep}%
  \BibitemOpen
  \bibfield  {author} {\bibinfo {author} {\bibfnamefont {P.~A.}\ \bibnamefont
  {McClarty}}\ and\ \bibinfo {author} {\bibfnamefont {J.~G.}\ \bibnamefont
  {Rau}},\ }\href {https://doi.org/10.1103/PhysRevB.100.100405} {\bibfield
  {journal} {\bibinfo  {journal} {Phys. Rev. B}\ }\textbf {\bibinfo {volume}
  {100}},\ \bibinfo {pages} {100405} (\bibinfo {year} {2019})}\BibitemShut
  {NoStop}%
\bibitem [{\citenamefont {Do}\ \emph {et~al.}(2022)\citenamefont {Do},
  \citenamefont {Kaneko}, \citenamefont {Kajimoto}, \citenamefont {Kamazawa},
  \citenamefont {Stone}, \citenamefont {Lin}, \citenamefont {Itoh},
  \citenamefont {Masuda}, \citenamefont {Samolyuk}, \citenamefont {Dagotto},
  \citenamefont {Meier}, \citenamefont {Sales}, \citenamefont {Miao},\ and\
  \citenamefont {Christianson}}]{FLQP_Hwan2022May}%
  \BibitemOpen
  \bibfield  {author} {\bibinfo {author} {\bibfnamefont {S.-H.}\ \bibnamefont
  {Do}}, \bibinfo {author} {\bibfnamefont {K.}~\bibnamefont {Kaneko}}, \bibinfo
  {author} {\bibfnamefont {R.}~\bibnamefont {Kajimoto}}, \bibinfo {author}
  {\bibfnamefont {K.}~\bibnamefont {Kamazawa}}, \bibinfo {author}
  {\bibfnamefont {M.~B.}\ \bibnamefont {Stone}}, \bibinfo {author}
  {\bibfnamefont {J.~Y.~Y.}\ \bibnamefont {Lin}}, \bibinfo {author}
  {\bibfnamefont {S.}~\bibnamefont {Itoh}}, \bibinfo {author} {\bibfnamefont
  {T.}~\bibnamefont {Masuda}}, \bibinfo {author} {\bibfnamefont {G.~D.}\
  \bibnamefont {Samolyuk}}, \bibinfo {author} {\bibfnamefont {E.}~\bibnamefont
  {Dagotto}}, \bibinfo {author} {\bibfnamefont {W.~R.}\ \bibnamefont {Meier}},
  \bibinfo {author} {\bibfnamefont {B.~C.}\ \bibnamefont {Sales}}, \bibinfo
  {author} {\bibfnamefont {H.}~\bibnamefont {Miao}},\ and\ \bibinfo {author}
  {\bibfnamefont {A.~D.}\ \bibnamefont {Christianson}},\ }\href
  {https://doi.org/10.1103/PhysRevB.105.L180403} {\bibfield  {journal}
  {\bibinfo  {journal} {Phys. Rev. B}\ }\textbf {\bibinfo {volume} {105}},\
  \bibinfo {pages} {L180403} (\bibinfo {year} {2022})}\BibitemShut {NoStop}%
\bibitem [{\citenamefont {Michen}\ and\ \citenamefont
  {Budich}(2022)}]{FLQP_Michen2022}%
  \BibitemOpen
  \bibfield  {author} {\bibinfo {author} {\bibfnamefont {B.}~\bibnamefont
  {Michen}}\ and\ \bibinfo {author} {\bibfnamefont {J.~C.}\ \bibnamefont
  {Budich}},\ }\href {https://doi.org/10.1103/PhysRevResearch.4.023248}
  {\bibfield  {journal} {\bibinfo  {journal} {Phys. Rev. Res.}\ }\textbf
  {\bibinfo {volume} {4}},\ \bibinfo {pages} {023248} (\bibinfo {year}
  {2022})}\BibitemShut {NoStop}%
\bibitem [{\citenamefont {Makris}\ \emph {et~al.}(2008)\citenamefont {Makris},
  \citenamefont {El-Ganainy}, \citenamefont {Christodoulides},\ and\
  \citenamefont {Musslimani}}]{NHOptical_Makris2008}%
  \BibitemOpen
  \bibfield  {author} {\bibinfo {author} {\bibfnamefont {K.~G.}\ \bibnamefont
  {Makris}}, \bibinfo {author} {\bibfnamefont {R.}~\bibnamefont {El-Ganainy}},
  \bibinfo {author} {\bibfnamefont {D.~N.}\ \bibnamefont {Christodoulides}},\
  and\ \bibinfo {author} {\bibfnamefont {Z.~H.}\ \bibnamefont {Musslimani}},\
  }\href {https://doi.org/10.1103/PhysRevLett.100.103904} {\bibfield  {journal}
  {\bibinfo  {journal} {Phys. Rev. Lett.}\ }\textbf {\bibinfo {volume} {100}},\
  \bibinfo {pages} {103904} (\bibinfo {year} {2008})}\BibitemShut {NoStop}%
\bibitem [{\citenamefont {Chong}\ \emph {et~al.}(2011)\citenamefont {Chong},
  \citenamefont {Ge},\ and\ \citenamefont {Stone}}]{NHOptical_Chong2011}%
  \BibitemOpen
  \bibfield  {author} {\bibinfo {author} {\bibfnamefont {Y.~D.}\ \bibnamefont
  {Chong}}, \bibinfo {author} {\bibfnamefont {L.}~\bibnamefont {Ge}},\ and\
  \bibinfo {author} {\bibfnamefont {A.~D.}\ \bibnamefont {Stone}},\ }\href
  {https://doi.org/10.1103/PhysRevLett.106.093902} {\bibfield  {journal}
  {\bibinfo  {journal} {Phys. Rev. Lett.}\ }\textbf {\bibinfo {volume} {106}},\
  \bibinfo {pages} {093902} (\bibinfo {year} {2011})}\BibitemShut {NoStop}%
\bibitem [{\citenamefont {Regensburger}\ \emph {et~al.}(2012)\citenamefont
  {Regensburger}, \citenamefont {Bersch}, \citenamefont {Miri}, \citenamefont
  {Onishchukov}, \citenamefont {Christodoulides},\ and\ \citenamefont
  {Peschel}}]{NHOptical_Regensburger2012}%
  \BibitemOpen
  \bibfield  {author} {\bibinfo {author} {\bibfnamefont {A.}~\bibnamefont
  {Regensburger}}, \bibinfo {author} {\bibfnamefont {C.}~\bibnamefont
  {Bersch}}, \bibinfo {author} {\bibfnamefont {M.-A.}\ \bibnamefont {Miri}},
  \bibinfo {author} {\bibfnamefont {G.}~\bibnamefont {Onishchukov}}, \bibinfo
  {author} {\bibfnamefont {D.~N.}\ \bibnamefont {Christodoulides}},\ and\
  \bibinfo {author} {\bibfnamefont {U.}~\bibnamefont {Peschel}},\ }\href
  {https://doi.org/10.1038/nature11298} {\bibfield  {journal} {\bibinfo
  {journal} {Nature}\ }\textbf {\bibinfo {volume} {488}},\ \bibinfo {pages}
  {167} (\bibinfo {year} {2012})}\BibitemShut {NoStop}%
\bibitem [{\citenamefont {Hodaei}\ \emph {et~al.}(2014)\citenamefont {Hodaei},
  \citenamefont {Miri}, \citenamefont {Heinrich}, \citenamefont
  {Christodoulides},\ and\ \citenamefont {Khajavikhan}}]{NHOptical_Hodaei2014}%
  \BibitemOpen
  \bibfield  {author} {\bibinfo {author} {\bibfnamefont {H.}~\bibnamefont
  {Hodaei}}, \bibinfo {author} {\bibfnamefont {M.-A.}\ \bibnamefont {Miri}},
  \bibinfo {author} {\bibfnamefont {M.}~\bibnamefont {Heinrich}}, \bibinfo
  {author} {\bibfnamefont {D.~N.}\ \bibnamefont {Christodoulides}},\ and\
  \bibinfo {author} {\bibfnamefont {M.}~\bibnamefont {Khajavikhan}},\ }\href
  {https://doi.org/10.1126/science.1258480} {\bibfield  {journal} {\bibinfo
  {journal} {Science}\ }\textbf {\bibinfo {volume} {346}},\ \bibinfo {pages}
  {975} (\bibinfo {year} {2014})}\BibitemShut {NoStop}%
\bibitem [{\citenamefont {Peng}\ \emph {et~al.}(2014)\citenamefont {Peng},
  \citenamefont {{\"O}zdemir}, \citenamefont {Lei}, \citenamefont {Monifi},
  \citenamefont {Gianfreda}, \citenamefont {Long}, \citenamefont {Fan},
  \citenamefont {Nori}, \citenamefont {Bender},\ and\ \citenamefont
  {Yang}}]{NHOptical_Peng2014}%
  \BibitemOpen
  \bibfield  {author} {\bibinfo {author} {\bibfnamefont {B.}~\bibnamefont
  {Peng}}, \bibinfo {author} {\bibfnamefont {{\c{S}}.~K.}\ \bibnamefont
  {{\"O}zdemir}}, \bibinfo {author} {\bibfnamefont {F.}~\bibnamefont {Lei}},
  \bibinfo {author} {\bibfnamefont {F.}~\bibnamefont {Monifi}}, \bibinfo
  {author} {\bibfnamefont {M.}~\bibnamefont {Gianfreda}}, \bibinfo {author}
  {\bibfnamefont {G.~L.}\ \bibnamefont {Long}}, \bibinfo {author}
  {\bibfnamefont {S.}~\bibnamefont {Fan}}, \bibinfo {author} {\bibfnamefont
  {F.}~\bibnamefont {Nori}}, \bibinfo {author} {\bibfnamefont {C.~M.}\
  \bibnamefont {Bender}},\ and\ \bibinfo {author} {\bibfnamefont
  {L.}~\bibnamefont {Yang}},\ }\href {https://doi.org/10.1038/nphys2927}
  {\bibfield  {journal} {\bibinfo  {journal} {Nature Physics}\ }\textbf
  {\bibinfo {volume} {10}},\ \bibinfo {pages} {394} (\bibinfo {year}
  {2014})}\BibitemShut {NoStop}%
\bibitem [{\citenamefont {Jing}\ \emph {et~al.}(2014)\citenamefont {Jing},
  \citenamefont {\"Ozdemir}, \citenamefont {L\"u}, \citenamefont {Zhang},
  \citenamefont {Yang},\ and\ \citenamefont {Nori}}]{NHOptical_Jing2014}%
  \BibitemOpen
  \bibfield  {author} {\bibinfo {author} {\bibfnamefont {H.}~\bibnamefont
  {Jing}}, \bibinfo {author} {\bibfnamefont {S.~K.}\ \bibnamefont {\"Ozdemir}},
  \bibinfo {author} {\bibfnamefont {X.-Y.}\ \bibnamefont {L\"u}}, \bibinfo
  {author} {\bibfnamefont {J.}~\bibnamefont {Zhang}}, \bibinfo {author}
  {\bibfnamefont {L.}~\bibnamefont {Yang}},\ and\ \bibinfo {author}
  {\bibfnamefont {F.}~\bibnamefont {Nori}},\ }\href
  {https://doi.org/10.1103/PhysRevLett.113.053604} {\bibfield  {journal}
  {\bibinfo  {journal} {Phys. Rev. Lett.}\ }\textbf {\bibinfo {volume} {113}},\
  \bibinfo {pages} {053604} (\bibinfo {year} {2014})}\BibitemShut {NoStop}%
\bibitem [{\citenamefont {Liu}\ \emph {et~al.}(2016)\citenamefont {Liu},
  \citenamefont {Zhang}, \citenamefont {{\"O}zdemir}, \citenamefont {Peng},
  \citenamefont {Jing}, \citenamefont {L\"u}, \citenamefont {Li}, \citenamefont
  {Yang}, \citenamefont {Nori},\ and\ \citenamefont {Liu}}]{NHOptical_Liu2016}%
  \BibitemOpen
  \bibfield  {author} {\bibinfo {author} {\bibfnamefont {Z.-P.}\ \bibnamefont
  {Liu}}, \bibinfo {author} {\bibfnamefont {J.}~\bibnamefont {Zhang}}, \bibinfo
  {author} {\bibfnamefont {{\c{S}}.~K.}\ \bibnamefont {{\"O}zdemir}}, \bibinfo
  {author} {\bibfnamefont {B.}~\bibnamefont {Peng}}, \bibinfo {author}
  {\bibfnamefont {H.}~\bibnamefont {Jing}}, \bibinfo {author} {\bibfnamefont
  {X.-Y.}\ \bibnamefont {L\"u}}, \bibinfo {author} {\bibfnamefont {C.-W.}\
  \bibnamefont {Li}}, \bibinfo {author} {\bibfnamefont {L.}~\bibnamefont
  {Yang}}, \bibinfo {author} {\bibfnamefont {F.}~\bibnamefont {Nori}},\ and\
  \bibinfo {author} {\bibfnamefont {Y.-X.}\ \bibnamefont {Liu}},\ }\href
  {https://doi.org/10.1103/PhysRevLett.117.110802} {\bibfield  {journal}
  {\bibinfo  {journal} {Phys. Rev. Lett.}\ }\textbf {\bibinfo {volume} {117}},\
  \bibinfo {pages} {110802} (\bibinfo {year} {2016})}\BibitemShut {NoStop}%
\bibitem [{\citenamefont {L\"u}\ \emph {et~al.}(2017)\citenamefont {L\"u},
  \citenamefont {\"Ozdemir}, \citenamefont {Kuang}, \citenamefont {Nori},\ and\
  \citenamefont {Jing}}]{NHOptical_Lu2017}%
  \BibitemOpen
  \bibfield  {author} {\bibinfo {author} {\bibfnamefont {H.}~\bibnamefont
  {L\"u}}, \bibinfo {author} {\bibfnamefont {S.~K.}\ \bibnamefont {\"Ozdemir}},
  \bibinfo {author} {\bibfnamefont {L.-M.}\ \bibnamefont {Kuang}}, \bibinfo
  {author} {\bibfnamefont {F.}~\bibnamefont {Nori}},\ and\ \bibinfo {author}
  {\bibfnamefont {H.}~\bibnamefont {Jing}},\ }\href
  {https://doi.org/10.1103/PhysRevApplied.8.044020} {\bibfield  {journal}
  {\bibinfo  {journal} {Phys. Rev. Appl.}\ }\textbf {\bibinfo {volume} {8}},\
  \bibinfo {pages} {044020} (\bibinfo {year} {2017})}\BibitemShut {NoStop}%
\bibitem [{\citenamefont {Soleymani}\ \emph {et~al.}(2022)\citenamefont
  {Soleymani}, \citenamefont {Zhong}, \citenamefont {Mokim}, \citenamefont
  {Rotter}, \citenamefont {El-Ganainy},\ and\ \citenamefont
  {{\"O}zdemir}}]{NHOptical_Soleymani2022}%
  \BibitemOpen
  \bibfield  {author} {\bibinfo {author} {\bibfnamefont {S.}~\bibnamefont
  {Soleymani}}, \bibinfo {author} {\bibfnamefont {Q.}~\bibnamefont {Zhong}},
  \bibinfo {author} {\bibfnamefont {M.}~\bibnamefont {Mokim}}, \bibinfo
  {author} {\bibfnamefont {S.}~\bibnamefont {Rotter}}, \bibinfo {author}
  {\bibfnamefont {R.}~\bibnamefont {El-Ganainy}},\ and\ \bibinfo {author}
  {\bibfnamefont {{\c{S}}.~K.}\ \bibnamefont {{\"O}zdemir}},\ }\href
  {https://doi.org/10.1038/s41467-022-27990-w} {\bibfield  {journal} {\bibinfo
  {journal} {Nature Communications}\ }\textbf {\bibinfo {volume} {13}},\
  \bibinfo {pages} {599} (\bibinfo {year} {2022})}\BibitemShut {NoStop}%
\bibitem [{\citenamefont {Zhang}\ \emph {et~al.}(2023)\citenamefont {Zhang},
  \citenamefont {Hu},\ and\ \citenamefont {Zhao}}]{NHOptical_Zhang2023Jan}%
  \BibitemOpen
  \bibfield  {author} {\bibinfo {author} {\bibfnamefont {X.}~\bibnamefont
  {Zhang}}, \bibinfo {author} {\bibfnamefont {J.}~\bibnamefont {Hu}},\ and\
  \bibinfo {author} {\bibfnamefont {N.}~\bibnamefont {Zhao}},\ }\href
  {https://doi.org/10.1103/PhysRevLett.130.023201} {\bibfield  {journal}
  {\bibinfo  {journal} {Phys. Rev. Lett.}\ }\textbf {\bibinfo {volume} {130}},\
  \bibinfo {pages} {023201} (\bibinfo {year} {2023})}\BibitemShut {NoStop}%
\bibitem [{\citenamefont {Arkhipov}\ \emph {et~al.}(2023)\citenamefont
  {Arkhipov}, \citenamefont {Miranowicz}, \citenamefont {Minganti},
  \citenamefont {{\"O}zdemir},\ and\ \citenamefont
  {Nori}}]{NHOptical_Arkhipov2023}%
  \BibitemOpen
  \bibfield  {author} {\bibinfo {author} {\bibfnamefont {I.~I.}\ \bibnamefont
  {Arkhipov}}, \bibinfo {author} {\bibfnamefont {A.}~\bibnamefont
  {Miranowicz}}, \bibinfo {author} {\bibfnamefont {F.}~\bibnamefont
  {Minganti}}, \bibinfo {author} {\bibfnamefont {{\c{S}}.~K.}\ \bibnamefont
  {{\"O}zdemir}},\ and\ \bibinfo {author} {\bibfnamefont {F.}~\bibnamefont
  {Nori}},\ }\href {https://doi.org/10.1038/s41467-023-37275-5} {\bibfield
  {journal} {\bibinfo  {journal} {Nature Communications}\ }\textbf {\bibinfo
  {volume} {14}},\ \bibinfo {pages} {2076} (\bibinfo {year}
  {2023})}\BibitemShut {NoStop}%
\bibitem [{\citenamefont {Helbig}\ \emph {et~al.}(2020)\citenamefont {Helbig},
  \citenamefont {Hofmann}, \citenamefont {Imhof}, \citenamefont {Abdelghany},
  \citenamefont {Kiessling}, \citenamefont {Molenkamp}, \citenamefont {Lee},
  \citenamefont {Szameit}, \citenamefont {Greiter},\ and\ \citenamefont
  {Thomale}}]{EC_Helbig2020}%
  \BibitemOpen
  \bibfield  {author} {\bibinfo {author} {\bibfnamefont {T.}~\bibnamefont
  {Helbig}}, \bibinfo {author} {\bibfnamefont {T.}~\bibnamefont {Hofmann}},
  \bibinfo {author} {\bibfnamefont {S.}~\bibnamefont {Imhof}}, \bibinfo
  {author} {\bibfnamefont {M.}~\bibnamefont {Abdelghany}}, \bibinfo {author}
  {\bibfnamefont {T.}~\bibnamefont {Kiessling}}, \bibinfo {author}
  {\bibfnamefont {L.~W.}\ \bibnamefont {Molenkamp}}, \bibinfo {author}
  {\bibfnamefont {C.~H.}\ \bibnamefont {Lee}}, \bibinfo {author} {\bibfnamefont
  {A.}~\bibnamefont {Szameit}}, \bibinfo {author} {\bibfnamefont
  {M.}~\bibnamefont {Greiter}},\ and\ \bibinfo {author} {\bibfnamefont
  {R.}~\bibnamefont {Thomale}},\ }\href
  {https://doi.org/10.1038/s41567-020-0922-9} {\bibfield  {journal} {\bibinfo
  {journal} {Nature Physics}\ }\textbf {\bibinfo {volume} {16}},\ \bibinfo
  {pages} {747} (\bibinfo {year} {2020})}\BibitemShut {NoStop}%
\bibitem [{\citenamefont {Chen}\ \emph {et~al.}(2023)\citenamefont {Chen},
  \citenamefont {Brand}, \citenamefont {Helbig}, \citenamefont {Hofmann},
  \citenamefont {Imhof}, \citenamefont {Fritzsche}, \citenamefont
  {Kie{\ss}ling}, \citenamefont {Stegmaier}, \citenamefont {Upreti},
  \citenamefont {Neupert}, \citenamefont {Bzdu{\v{s}}ek}, \citenamefont
  {Greiter}, \citenamefont {Thomale},\ and\ \citenamefont
  {Boettcher}}]{EC_Chen2023}%
  \BibitemOpen
  \bibfield  {author} {\bibinfo {author} {\bibfnamefont {A.}~\bibnamefont
  {Chen}}, \bibinfo {author} {\bibfnamefont {H.}~\bibnamefont {Brand}},
  \bibinfo {author} {\bibfnamefont {T.}~\bibnamefont {Helbig}}, \bibinfo
  {author} {\bibfnamefont {T.}~\bibnamefont {Hofmann}}, \bibinfo {author}
  {\bibfnamefont {S.}~\bibnamefont {Imhof}}, \bibinfo {author} {\bibfnamefont
  {A.}~\bibnamefont {Fritzsche}}, \bibinfo {author} {\bibfnamefont
  {T.}~\bibnamefont {Kie{\ss}ling}}, \bibinfo {author} {\bibfnamefont
  {A.}~\bibnamefont {Stegmaier}}, \bibinfo {author} {\bibfnamefont {L.~K.}\
  \bibnamefont {Upreti}}, \bibinfo {author} {\bibfnamefont {T.}~\bibnamefont
  {Neupert}}, \bibinfo {author} {\bibfnamefont {T.}~\bibnamefont
  {Bzdu{\v{s}}ek}}, \bibinfo {author} {\bibfnamefont {M.}~\bibnamefont
  {Greiter}}, \bibinfo {author} {\bibfnamefont {R.}~\bibnamefont {Thomale}},\
  and\ \bibinfo {author} {\bibfnamefont {I.}~\bibnamefont {Boettcher}},\ }\href
  {https://doi.org/10.1038/s41467-023-36359-6} {\bibfield  {journal} {\bibinfo
  {journal} {Nature Communications}\ }\textbf {\bibinfo {volume} {14}},\
  \bibinfo {pages} {622} (\bibinfo {year} {2023})}\BibitemShut {NoStop}%
\bibitem [{\citenamefont {Wu}\ \emph {et~al.}(2023)\citenamefont {Wu},
  \citenamefont {Zhao}, \citenamefont {Kang}, \citenamefont {Weng},
  \citenamefont {Chi}, \citenamefont {Peng}, \citenamefont {Liu}, \citenamefont
  {Werner}, \citenamefont {Meng},\ and\ \citenamefont {Zhou}}]{EC_Maopeng2023}%
  \BibitemOpen
  \bibfield  {author} {\bibinfo {author} {\bibfnamefont {M.}~\bibnamefont
  {Wu}}, \bibinfo {author} {\bibfnamefont {Q.}~\bibnamefont {Zhao}}, \bibinfo
  {author} {\bibfnamefont {L.}~\bibnamefont {Kang}}, \bibinfo {author}
  {\bibfnamefont {M.}~\bibnamefont {Weng}}, \bibinfo {author} {\bibfnamefont
  {Z.}~\bibnamefont {Chi}}, \bibinfo {author} {\bibfnamefont {R.}~\bibnamefont
  {Peng}}, \bibinfo {author} {\bibfnamefont {J.}~\bibnamefont {Liu}}, \bibinfo
  {author} {\bibfnamefont {D.~H.}\ \bibnamefont {Werner}}, \bibinfo {author}
  {\bibfnamefont {Y.}~\bibnamefont {Meng}},\ and\ \bibinfo {author}
  {\bibfnamefont {J.}~\bibnamefont {Zhou}},\ }\href
  {https://doi.org/10.1103/PhysRevB.107.064307} {\bibfield  {journal} {\bibinfo
   {journal} {Phys. Rev. B}\ }\textbf {\bibinfo {volume} {107}},\ \bibinfo
  {pages} {064307} (\bibinfo {year} {2023})}\BibitemShut {NoStop}%
\bibitem [{\citenamefont {Nelson}\ and\ \citenamefont
  {Shnerb}(1998)}]{Bio_Nelson1998}%
  \BibitemOpen
  \bibfield  {author} {\bibinfo {author} {\bibfnamefont {D.~R.}\ \bibnamefont
  {Nelson}}\ and\ \bibinfo {author} {\bibfnamefont {N.~M.}\ \bibnamefont
  {Shnerb}},\ }\href {https://doi.org/10.1103/PhysRevE.58.1383} {\bibfield
  {journal} {\bibinfo  {journal} {Phys. Rev. E}\ }\textbf {\bibinfo {volume}
  {58}},\ \bibinfo {pages} {1383} (\bibinfo {year} {1998})}\BibitemShut
  {NoStop}%
\bibitem [{\citenamefont {Amir}\ \emph {et~al.}(2016)\citenamefont {Amir},
  \citenamefont {Hatano},\ and\ \citenamefont {Nelson}}]{Bio_Amir2016}%
  \BibitemOpen
  \bibfield  {author} {\bibinfo {author} {\bibfnamefont {A.}~\bibnamefont
  {Amir}}, \bibinfo {author} {\bibfnamefont {N.}~\bibnamefont {Hatano}},\ and\
  \bibinfo {author} {\bibfnamefont {D.~R.}\ \bibnamefont {Nelson}},\ }\href
  {https://doi.org/10.1103/PhysRevE.93.042310} {\bibfield  {journal} {\bibinfo
  {journal} {Phys. Rev. E}\ }\textbf {\bibinfo {volume} {93}},\ \bibinfo
  {pages} {042310} (\bibinfo {year} {2016})}\BibitemShut {NoStop}%
\bibitem [{\citenamefont {Murugan}\ and\ \citenamefont
  {Vaikuntanathan}(2017)}]{Bio_Murugan2017}%
  \BibitemOpen
  \bibfield  {author} {\bibinfo {author} {\bibfnamefont {A.}~\bibnamefont
  {Murugan}}\ and\ \bibinfo {author} {\bibfnamefont {S.}~\bibnamefont
  {Vaikuntanathan}},\ }\href {https://doi.org/10.1038/ncomms13881} {\bibfield
  {journal} {\bibinfo  {journal} {Nature Communications}\ }\textbf {\bibinfo
  {volume} {8}},\ \bibinfo {pages} {13881} (\bibinfo {year}
  {2017})}\BibitemShut {NoStop}%
\bibitem [{\citenamefont {Bender}(2007)}]{EPs_Bender2007}%
  \BibitemOpen
  \bibfield  {author} {\bibinfo {author} {\bibfnamefont {C.~M.}\ \bibnamefont
  {Bender}},\ }\href {https://doi.org/10.1088/0034-4885/70/6/R03} {\bibfield
  {journal} {\bibinfo  {journal} {Reports on Progress in Physics}\ }\textbf
  {\bibinfo {volume} {70}},\ \bibinfo {pages} {947} (\bibinfo {year}
  {2007})}\BibitemShut {NoStop}%
\bibitem [{\citenamefont {Heiss}(2012)}]{EPs_Heiss2012}%
  \BibitemOpen
  \bibfield  {author} {\bibinfo {author} {\bibfnamefont {W.~D.}\ \bibnamefont
  {Heiss}},\ }\href {https://doi.org/10.1088/1751-8113/45/44/444016} {\bibfield
   {journal} {\bibinfo  {journal} {Journal of Physics A: Mathematical and
  Theoretical}\ }\textbf {\bibinfo {volume} {45}},\ \bibinfo {pages} {444016}
  (\bibinfo {year} {2012})}\BibitemShut {NoStop}%
\bibitem [{\citenamefont {Yao}\ and\ \citenamefont
  {Wang}(2018)}]{BOBBC_Yao2018}%
  \BibitemOpen
  \bibfield  {author} {\bibinfo {author} {\bibfnamefont {S.}~\bibnamefont
  {Yao}}\ and\ \bibinfo {author} {\bibfnamefont {Z.}~\bibnamefont {Wang}},\
  }\href {https://doi.org/10.1103/PhysRevLett.121.086803} {\bibfield  {journal}
  {\bibinfo  {journal} {Phys. Rev. Lett.}\ }\textbf {\bibinfo {volume} {121}},\
  \bibinfo {pages} {086803} (\bibinfo {year} {2018})}\BibitemShut {NoStop}%
\bibitem [{\citenamefont {Kunst}\ \emph {et~al.}(2018)\citenamefont {Kunst},
  \citenamefont {Edvardsson}, \citenamefont {Budich},\ and\ \citenamefont
  {Bergholtz}}]{BOBBC_Kunst2018}%
  \BibitemOpen
  \bibfield  {author} {\bibinfo {author} {\bibfnamefont {F.~K.}\ \bibnamefont
  {Kunst}}, \bibinfo {author} {\bibfnamefont {E.}~\bibnamefont {Edvardsson}},
  \bibinfo {author} {\bibfnamefont {J.~C.}\ \bibnamefont {Budich}},\ and\
  \bibinfo {author} {\bibfnamefont {E.~J.}\ \bibnamefont {Bergholtz}},\ }\href
  {https://doi.org/10.1103/PhysRevLett.121.026808} {\bibfield  {journal}
  {\bibinfo  {journal} {Phys. Rev. Lett.}\ }\textbf {\bibinfo {volume} {121}},\
  \bibinfo {pages} {026808} (\bibinfo {year} {2018})}\BibitemShut {NoStop}%
\bibitem [{\citenamefont {Lee}(2016)}]{Skin_Lee2016}%
  \BibitemOpen
  \bibfield  {author} {\bibinfo {author} {\bibfnamefont {T.~E.}\ \bibnamefont
  {Lee}},\ }\href {https://doi.org/10.1103/PhysRevLett.116.133903} {\bibfield
  {journal} {\bibinfo  {journal} {Phys. Rev. Lett.}\ }\textbf {\bibinfo
  {volume} {116}},\ \bibinfo {pages} {133903} (\bibinfo {year}
  {2016})}\BibitemShut {NoStop}%
\bibitem [{\citenamefont {Xiong}(2018)}]{Skin_Xiong2018}%
  \BibitemOpen
  \bibfield  {author} {\bibinfo {author} {\bibfnamefont {Y.}~\bibnamefont
  {Xiong}},\ }\href {https://doi.org/10.1088/2399-6528/aab64a} {\bibfield
  {journal} {\bibinfo  {journal} {Journal of Physics Communications}\ }\textbf
  {\bibinfo {volume} {2}},\ \bibinfo {pages} {035043} (\bibinfo {year}
  {2018})}\BibitemShut {NoStop}%
\bibitem [{\citenamefont {Yao}\ \emph {et~al.}(2018)\citenamefont {Yao},
  \citenamefont {Song},\ and\ \citenamefont {Wang}}]{Skin_Yao2018}%
  \BibitemOpen
  \bibfield  {author} {\bibinfo {author} {\bibfnamefont {S.}~\bibnamefont
  {Yao}}, \bibinfo {author} {\bibfnamefont {F.}~\bibnamefont {Song}},\ and\
  \bibinfo {author} {\bibfnamefont {Z.}~\bibnamefont {Wang}},\ }\href
  {https://doi.org/10.1103/PhysRevLett.121.136802} {\bibfield  {journal}
  {\bibinfo  {journal} {Phys. Rev. Lett.}\ }\textbf {\bibinfo {volume} {121}},\
  \bibinfo {pages} {136802} (\bibinfo {year} {2018})}\BibitemShut {NoStop}%
\bibitem [{\citenamefont {Bernard}\ and\ \citenamefont
  {LeClair}(2002)}]{Bernard2002}%
  \BibitemOpen
  \bibfield  {author} {\bibinfo {author} {\bibfnamefont {D.}~\bibnamefont
  {Bernard}}\ and\ \bibinfo {author} {\bibfnamefont {A.}~\bibnamefont
  {LeClair}}\ }(\bibinfo  {publisher} {Springer Netherlands},\ \bibinfo
  {address} {Dordrecht},\ \bibinfo {year} {2002})\ pp.\ \bibinfo {pages}
  {207--214}\BibitemShut {NoStop}%
\bibitem [{\citenamefont {Zhou}\ and\ \citenamefont
  {Lee}(2019)}]{HZhou_Table_nonHermitianSymmetries}%
  \BibitemOpen
  \bibfield  {author} {\bibinfo {author} {\bibfnamefont {H.}~\bibnamefont
  {Zhou}}\ and\ \bibinfo {author} {\bibfnamefont {J.~Y.}\ \bibnamefont {Lee}},\
  }\href {https://doi.org/10.1103/PhysRevB.99.235112} {\bibfield  {journal}
  {\bibinfo  {journal} {Phys. Rev. B}\ }\textbf {\bibinfo {volume} {99}},\
  \bibinfo {pages} {235112} (\bibinfo {year} {2019})}\BibitemShut {NoStop}%
\bibitem [{\citenamefont {Kawabata}\ \emph
  {et~al.}(2019{\natexlab{b}})\citenamefont {Kawabata}, \citenamefont
  {Shiozaki}, \citenamefont {Ueda},\ and\ \citenamefont
  {Sato}}]{KKawabata_prx_nonHermitianSymm}%
  \BibitemOpen
  \bibfield  {author} {\bibinfo {author} {\bibfnamefont {K.}~\bibnamefont
  {Kawabata}}, \bibinfo {author} {\bibfnamefont {K.}~\bibnamefont {Shiozaki}},
  \bibinfo {author} {\bibfnamefont {M.}~\bibnamefont {Ueda}},\ and\ \bibinfo
  {author} {\bibfnamefont {M.}~\bibnamefont {Sato}},\ }\href
  {https://doi.org/10.1103/PhysRevX.9.041015} {\bibfield  {journal} {\bibinfo
  {journal} {Phys. Rev. X}\ }\textbf {\bibinfo {volume} {9}},\ \bibinfo {pages}
  {041015} (\bibinfo {year} {2019}{\natexlab{b}})}\BibitemShut {NoStop}%
\bibitem [{\citenamefont {Kawabata}\ \emph
  {et~al.}(2019{\natexlab{c}})\citenamefont {Kawabata}, \citenamefont
  {Higashikawa}, \citenamefont {Gong}, \citenamefont {Ashida},\ and\
  \citenamefont {Ueda}}]{KKawabata_Unification_TRPT}%
  \BibitemOpen
  \bibfield  {author} {\bibinfo {author} {\bibfnamefont {K.}~\bibnamefont
  {Kawabata}}, \bibinfo {author} {\bibfnamefont {S.}~\bibnamefont
  {Higashikawa}}, \bibinfo {author} {\bibfnamefont {Z.}~\bibnamefont {Gong}},
  \bibinfo {author} {\bibfnamefont {Y.}~\bibnamefont {Ashida}},\ and\ \bibinfo
  {author} {\bibfnamefont {M.}~\bibnamefont {Ueda}},\ }\href
  {https://doi.org/10.1038/s41467-018-08254-y} {\bibfield  {journal} {\bibinfo
  {journal} {Nature Communications}\ }\textbf {\bibinfo {volume} {10}},\
  \bibinfo {pages} {297} (\bibinfo {year} {2019}{\natexlab{c}})}\BibitemShut
  {NoStop}%
\bibitem [{\citenamefont {Esaki}\ \emph {et~al.}(2011)\citenamefont {Esaki},
  \citenamefont {Sato}, \citenamefont {Hasebe},\ and\ \citenamefont
  {Kohmoto}}]{KentaEsaki_Edge_states_TR}%
  \BibitemOpen
  \bibfield  {author} {\bibinfo {author} {\bibfnamefont {K.}~\bibnamefont
  {Esaki}}, \bibinfo {author} {\bibfnamefont {M.}~\bibnamefont {Sato}},
  \bibinfo {author} {\bibfnamefont {K.}~\bibnamefont {Hasebe}},\ and\ \bibinfo
  {author} {\bibfnamefont {M.}~\bibnamefont {Kohmoto}},\ }\href
  {https://doi.org/10.1103/PhysRevB.84.205128} {\bibfield  {journal} {\bibinfo
  {journal} {Phys. Rev. B}\ }\textbf {\bibinfo {volume} {84}},\ \bibinfo
  {pages} {205128} (\bibinfo {year} {2011})}\BibitemShut {NoStop}%
\bibitem [{\citenamefont {Budich}\ \emph {et~al.}(2019)\citenamefont {Budich},
  \citenamefont {Carlstr\"om}, \citenamefont {Kunst},\ and\ \citenamefont
  {Bergholtz}}]{SymmClassification_Budich2019}%
  \BibitemOpen
  \bibfield  {author} {\bibinfo {author} {\bibfnamefont {J.~C.}\ \bibnamefont
  {Budich}}, \bibinfo {author} {\bibfnamefont {J.}~\bibnamefont {Carlstr\"om}},
  \bibinfo {author} {\bibfnamefont {F.~K.}\ \bibnamefont {Kunst}},\ and\
  \bibinfo {author} {\bibfnamefont {E.~J.}\ \bibnamefont {Bergholtz}},\ }\href
  {https://doi.org/10.1103/PhysRevB.99.041406} {\bibfield  {journal} {\bibinfo
  {journal} {Phys. Rev. B}\ }\textbf {\bibinfo {volume} {99}},\ \bibinfo
  {pages} {041406} (\bibinfo {year} {2019})}\BibitemShut {NoStop}%
\bibitem [{\citenamefont {Zhang}\ \emph {et~al.}(2013)\citenamefont {Zhang},
  \citenamefont {Kane},\ and\ \citenamefont {Mele}}]{HOTI_Zhang2013}%
  \BibitemOpen
  \bibfield  {author} {\bibinfo {author} {\bibfnamefont {F.}~\bibnamefont
  {Zhang}}, \bibinfo {author} {\bibfnamefont {C.~L.}\ \bibnamefont {Kane}},\
  and\ \bibinfo {author} {\bibfnamefont {E.~J.}\ \bibnamefont {Mele}},\ }\href
  {https://doi.org/10.1103/PhysRevLett.110.046404} {\bibfield  {journal}
  {\bibinfo  {journal} {Phys. Rev. Lett.}\ }\textbf {\bibinfo {volume} {110}},\
  \bibinfo {pages} {046404} (\bibinfo {year} {2013})}\BibitemShut {NoStop}%
\bibitem [{\citenamefont {Benalcazar}\ \emph
  {et~al.}(2017{\natexlab{a}})\citenamefont {Benalcazar}, \citenamefont
  {Bernevig},\ and\ \citenamefont {Hughes}}]{HOTI_Benalcazar2017}%
  \BibitemOpen
  \bibfield  {author} {\bibinfo {author} {\bibfnamefont {W.~A.}\ \bibnamefont
  {Benalcazar}}, \bibinfo {author} {\bibfnamefont {B.~A.}\ \bibnamefont
  {Bernevig}},\ and\ \bibinfo {author} {\bibfnamefont {T.~L.}\ \bibnamefont
  {Hughes}},\ }\href {https://doi.org/10.1126/science.aah6442} {\bibfield
  {journal} {\bibinfo  {journal} {Science}\ }\textbf {\bibinfo {volume}
  {357}},\ \bibinfo {pages} {61} (\bibinfo {year}
  {2017}{\natexlab{a}})}\BibitemShut {NoStop}%
\bibitem [{\citenamefont {Benalcazar}\ \emph
  {et~al.}(2017{\natexlab{b}})\citenamefont {Benalcazar}, \citenamefont
  {Bernevig},\ and\ \citenamefont {Hughes}}]{HOTI_Benalcazar2017Dec}%
  \BibitemOpen
  \bibfield  {author} {\bibinfo {author} {\bibfnamefont {W.~A.}\ \bibnamefont
  {Benalcazar}}, \bibinfo {author} {\bibfnamefont {B.~A.}\ \bibnamefont
  {Bernevig}},\ and\ \bibinfo {author} {\bibfnamefont {T.~L.}\ \bibnamefont
  {Hughes}},\ }\href {https://doi.org/10.1103/PhysRevB.96.245115} {\bibfield
  {journal} {\bibinfo  {journal} {Phys. Rev. B}\ }\textbf {\bibinfo {volume}
  {96}},\ \bibinfo {pages} {245115} (\bibinfo {year}
  {2017}{\natexlab{b}})}\BibitemShut {NoStop}%
\bibitem [{\citenamefont {Langbehn}\ \emph {et~al.}(2017)\citenamefont
  {Langbehn}, \citenamefont {Peng}, \citenamefont {Trifunovic}, \citenamefont
  {von Oppen},\ and\ \citenamefont {Brouwer}}]{HOTI_Josias2017}%
  \BibitemOpen
  \bibfield  {author} {\bibinfo {author} {\bibfnamefont {J.}~\bibnamefont
  {Langbehn}}, \bibinfo {author} {\bibfnamefont {Y.}~\bibnamefont {Peng}},
  \bibinfo {author} {\bibfnamefont {L.}~\bibnamefont {Trifunovic}}, \bibinfo
  {author} {\bibfnamefont {F.}~\bibnamefont {von Oppen}},\ and\ \bibinfo
  {author} {\bibfnamefont {P.~W.}\ \bibnamefont {Brouwer}},\ }\href
  {https://doi.org/10.1103/PhysRevLett.119.246401} {\bibfield  {journal}
  {\bibinfo  {journal} {Phys. Rev. Lett.}\ }\textbf {\bibinfo {volume} {119}},\
  \bibinfo {pages} {246401} (\bibinfo {year} {2017})}\BibitemShut {NoStop}%
\bibitem [{\citenamefont {Schindler}\ \emph {et~al.}(2018)\citenamefont
  {Schindler}, \citenamefont {Cook}, \citenamefont {Vergniory}, \citenamefont
  {Wang}, \citenamefont {Parkin}, \citenamefont {Bernevig},\ and\ \citenamefont
  {Neupert}}]{HOTI_Frank2018}%
  \BibitemOpen
  \bibfield  {author} {\bibinfo {author} {\bibfnamefont {F.}~\bibnamefont
  {Schindler}}, \bibinfo {author} {\bibfnamefont {A.~M.}\ \bibnamefont {Cook}},
  \bibinfo {author} {\bibfnamefont {M.~G.}\ \bibnamefont {Vergniory}}, \bibinfo
  {author} {\bibfnamefont {Z.}~\bibnamefont {Wang}}, \bibinfo {author}
  {\bibfnamefont {S.~S.~P.}\ \bibnamefont {Parkin}}, \bibinfo {author}
  {\bibfnamefont {B.~A.}\ \bibnamefont {Bernevig}},\ and\ \bibinfo {author}
  {\bibfnamefont {T.}~\bibnamefont {Neupert}},\ }\href
  {https://doi.org/10.1126/sciadv.aat0346} {\bibfield  {journal} {\bibinfo
  {journal} {Science Advances}\ }\textbf {\bibinfo {volume} {4}},\ \bibinfo
  {pages} {eaat0346} (\bibinfo {year} {2018})}\BibitemShut {NoStop}%
\bibitem [{\citenamefont {Xie}\ \emph {et~al.}(2021)\citenamefont {Xie},
  \citenamefont {Wang}, \citenamefont {Zhang}, \citenamefont {Zhan},
  \citenamefont {Jiang}, \citenamefont {Lu},\ and\ \citenamefont
  {Chen}}]{HOTI_Xie2021}%
  \BibitemOpen
  \bibfield  {author} {\bibinfo {author} {\bibfnamefont {B.}~\bibnamefont
  {Xie}}, \bibinfo {author} {\bibfnamefont {H.-X.}\ \bibnamefont {Wang}},
  \bibinfo {author} {\bibfnamefont {X.}~\bibnamefont {Zhang}}, \bibinfo
  {author} {\bibfnamefont {P.}~\bibnamefont {Zhan}}, \bibinfo {author}
  {\bibfnamefont {J.-H.}\ \bibnamefont {Jiang}}, \bibinfo {author}
  {\bibfnamefont {M.}~\bibnamefont {Lu}},\ and\ \bibinfo {author}
  {\bibfnamefont {Y.}~\bibnamefont {Chen}},\ }\href
  {https://doi.org/10.1038/s42254-021-00323-4} {\bibfield  {journal} {\bibinfo
  {journal} {Nature Reviews Physics}\ }\textbf {\bibinfo {volume} {3}},\
  \bibinfo {pages} {520} (\bibinfo {year} {2021})}\BibitemShut {NoStop}%
\bibitem [{\citenamefont {Varjas}\ \emph {et~al.}(2019)\citenamefont {Varjas},
  \citenamefont {Lau}, \citenamefont {P\"oyh\"onen}, \citenamefont {Akhmerov},
  \citenamefont {Pikulin},\ and\ \citenamefont {Fulga}}]{Daniel-Varjas}%
  \BibitemOpen
  \bibfield  {author} {\bibinfo {author} {\bibfnamefont {D.}~\bibnamefont
  {Varjas}}, \bibinfo {author} {\bibfnamefont {A.}~\bibnamefont {Lau}},
  \bibinfo {author} {\bibfnamefont {K.}~\bibnamefont {P\"oyh\"onen}}, \bibinfo
  {author} {\bibfnamefont {A.~R.}\ \bibnamefont {Akhmerov}}, \bibinfo {author}
  {\bibfnamefont {D.~I.}\ \bibnamefont {Pikulin}},\ and\ \bibinfo {author}
  {\bibfnamefont {I.~C.}\ \bibnamefont {Fulga}},\ }\href
  {https://doi.org/10.1103/PhysRevLett.123.196401} {\bibfield  {journal}
  {\bibinfo  {journal} {Phys. Rev. Lett.}\ }\textbf {\bibinfo {volume} {123}},\
  \bibinfo {pages} {196401} (\bibinfo {year} {2019})}\BibitemShut {NoStop}%
\bibitem [{\citenamefont {Chen}\ \emph {et~al.}(2020)\citenamefont {Chen},
  \citenamefont {Chen}, \citenamefont {Gao}, \citenamefont {Zhou},\ and\
  \citenamefont {Xu}}]{RuiChen_etal_prl}%
  \BibitemOpen
  \bibfield  {author} {\bibinfo {author} {\bibfnamefont {R.}~\bibnamefont
  {Chen}}, \bibinfo {author} {\bibfnamefont {C.-Z.}\ \bibnamefont {Chen}},
  \bibinfo {author} {\bibfnamefont {J.-H.}\ \bibnamefont {Gao}}, \bibinfo
  {author} {\bibfnamefont {B.}~\bibnamefont {Zhou}},\ and\ \bibinfo {author}
  {\bibfnamefont {D.-H.}\ \bibnamefont {Xu}},\ }\href
  {https://doi.org/10.1103/PhysRevLett.124.036803} {\bibfield  {journal}
  {\bibinfo  {journal} {Phys. Rev. Lett.}\ }\textbf {\bibinfo {volume} {124}},\
  \bibinfo {pages} {036803} (\bibinfo {year} {2020})}\BibitemShut {NoStop}%
\bibitem [{\citenamefont {Agarwala}\ \emph {et~al.}(2020)\citenamefont
  {Agarwala}, \citenamefont {Juri\ifmmode \check{c}\else
  \v{c}\fi{}i\ifmmode~\acute{c}\else \'{c}\fi{}},\ and\ \citenamefont
  {Roy}}]{AAgarwala_prr_HOTIAmorphou_}%
  \BibitemOpen
  \bibfield  {author} {\bibinfo {author} {\bibfnamefont {A.}~\bibnamefont
  {Agarwala}}, \bibinfo {author} {\bibfnamefont {V.}~\bibnamefont {Juri\ifmmode
  \check{c}\else \v{c}\fi{}i\ifmmode~\acute{c}\else \'{c}\fi{}}},\ and\
  \bibinfo {author} {\bibfnamefont {B.}~\bibnamefont {Roy}},\ }\href
  {https://doi.org/10.1103/PhysRevResearch.2.012067} {\bibfield  {journal}
  {\bibinfo  {journal} {Phys. Rev. Res.}\ }\textbf {\bibinfo {volume} {2}},\
  \bibinfo {pages} {012067} (\bibinfo {year} {2020})}\BibitemShut {NoStop}%
\bibitem [{\citenamefont {Liu}\ \emph {et~al.}(2019)\citenamefont {Liu},
  \citenamefont {Zhang}, \citenamefont {Ai}, \citenamefont {Gong},
  \citenamefont {Kawabata}, \citenamefont {Ueda},\ and\ \citenamefont
  {Nori}}]{Liu-Tao}%
  \BibitemOpen
  \bibfield  {author} {\bibinfo {author} {\bibfnamefont {T.}~\bibnamefont
  {Liu}}, \bibinfo {author} {\bibfnamefont {Y.-R.}\ \bibnamefont {Zhang}},
  \bibinfo {author} {\bibfnamefont {Q.}~\bibnamefont {Ai}}, \bibinfo {author}
  {\bibfnamefont {Z.}~\bibnamefont {Gong}}, \bibinfo {author} {\bibfnamefont
  {K.}~\bibnamefont {Kawabata}}, \bibinfo {author} {\bibfnamefont
  {M.}~\bibnamefont {Ueda}},\ and\ \bibinfo {author} {\bibfnamefont
  {F.}~\bibnamefont {Nori}},\ }\href
  {https://doi.org/10.1103/PhysRevLett.122.076801} {\bibfield  {journal}
  {\bibinfo  {journal} {Phys. Rev. Lett.}\ }\textbf {\bibinfo {volume} {122}},\
  \bibinfo {pages} {076801} (\bibinfo {year} {2019})}\BibitemShut {NoStop}%
\bibitem [{\citenamefont {Kawabata}\ and\ \citenamefont
  {Sato}(2020)}]{KKawabata_prr_realspectra}%
  \BibitemOpen
  \bibfield  {author} {\bibinfo {author} {\bibfnamefont {K.}~\bibnamefont
  {Kawabata}}\ and\ \bibinfo {author} {\bibfnamefont {M.}~\bibnamefont
  {Sato}},\ }\href {https://doi.org/10.1103/PhysRevResearch.2.033391}
  {\bibfield  {journal} {\bibinfo  {journal} {Phys. Rev. Res.}\ }\textbf
  {\bibinfo {volume} {2}},\ \bibinfo {pages} {033391} (\bibinfo {year}
  {2020})}\BibitemShut {NoStop}%
\bibitem [{\citenamefont {Agarwala}\ and\ \citenamefont
  {Shenoy}(2017)}]{AAgarwala_TI_Amorphous}%
  \BibitemOpen
  \bibfield  {author} {\bibinfo {author} {\bibfnamefont {A.}~\bibnamefont
  {Agarwala}}\ and\ \bibinfo {author} {\bibfnamefont {V.~B.}\ \bibnamefont
  {Shenoy}},\ }\href {https://doi.org/10.1103/PhysRevLett.118.236402}
  {\bibfield  {journal} {\bibinfo  {journal} {Phys. Rev. Lett.}\ }\textbf
  {\bibinfo {volume} {118}},\ \bibinfo {pages} {236402} (\bibinfo {year}
  {2017})}\BibitemShut {NoStop}%
\bibitem [{\citenamefont {Mitchell}\ \emph {et~al.}(2018)\citenamefont
  {Mitchell}, \citenamefont {Nash}, \citenamefont {Hexner}, \citenamefont
  {Turner},\ and\ \citenamefont {Irvine}}]{Mitchell2018_Amorphou_TI}%
  \BibitemOpen
  \bibfield  {author} {\bibinfo {author} {\bibfnamefont {N.~P.}\ \bibnamefont
  {Mitchell}}, \bibinfo {author} {\bibfnamefont {L.~M.}\ \bibnamefont {Nash}},
  \bibinfo {author} {\bibfnamefont {D.}~\bibnamefont {Hexner}}, \bibinfo
  {author} {\bibfnamefont {A.~M.}\ \bibnamefont {Turner}},\ and\ \bibinfo
  {author} {\bibfnamefont {W.~T.~M.}\ \bibnamefont {Irvine}},\ }\href
  {https://doi.org/10.1038/s41567-017-0024-5} {\bibfield  {journal} {\bibinfo
  {journal} {Nature Physics}\ }\textbf {\bibinfo {volume} {14}},\ \bibinfo
  {pages} {380} (\bibinfo {year} {2018})}\BibitemShut {NoStop}%
\bibitem [{\citenamefont
  {Mostafazadeh}(2002{\natexlab{a}})}]{AliM_pseudo_antihermiticity}%
  \BibitemOpen
  \bibfield  {author} {\bibinfo {author} {\bibfnamefont {A.}~\bibnamefont
  {Mostafazadeh}},\ }\href {https://doi.org/10.1063/1.1418246} {\bibfield
  {journal} {\bibinfo  {journal} {Journal of Mathematical Physics}\ }\textbf
  {\bibinfo {volume} {43}},\ \bibinfo {pages} {205} (\bibinfo {year}
  {2002}{\natexlab{a}})}\BibitemShut {NoStop}%
\bibitem [{\citenamefont
  {Mostafazadeh}(2002{\natexlab{b}})}]{AliM_pseudo_antihermiticityII}%
  \BibitemOpen
  \bibfield  {author} {\bibinfo {author} {\bibfnamefont {A.}~\bibnamefont
  {Mostafazadeh}},\ }\href {https://doi.org/10.1063/1.1461427} {\bibfield
  {journal} {\bibinfo  {journal} {Journal of Mathematical Physics}\ }\textbf
  {\bibinfo {volume} {43}},\ \bibinfo {pages} {2814} (\bibinfo {year}
  {2002}{\natexlab{b}})}\BibitemShut {NoStop}%
\bibitem [{\citenamefont
  {Mostafazadeh}(2002{\natexlab{c}})}]{AliM_pseudo_antihermiticityIII}%
  \BibitemOpen
  \bibfield  {author} {\bibinfo {author} {\bibfnamefont {A.}~\bibnamefont
  {Mostafazadeh}},\ }\href {https://doi.org/10.1063/1.1489072} {\bibfield
  {journal} {\bibinfo  {journal} {Journal of Mathematical Physics}\ }\textbf
  {\bibinfo {volume} {43}},\ \bibinfo {pages} {3944} (\bibinfo {year}
  {2002}{\natexlab{c}})}\BibitemShut {NoStop}%
\bibitem [{\citenamefont {Sato}\ \emph {et~al.}(2012)\citenamefont {Sato},
  \citenamefont {Hasebe}, \citenamefont {Esaki},\ and\ \citenamefont
  {Kohmoto}}]{kramers_deg_Sato}%
  \BibitemOpen
  \bibfield  {author} {\bibinfo {author} {\bibfnamefont {M.}~\bibnamefont
  {Sato}}, \bibinfo {author} {\bibfnamefont {K.}~\bibnamefont {Hasebe}},
  \bibinfo {author} {\bibfnamefont {K.}~\bibnamefont {Esaki}},\ and\ \bibinfo
  {author} {\bibfnamefont {M.}~\bibnamefont {Kohmoto}},\ }\href
  {https://doi.org/10.1143/PTP.127.937} {\bibfield  {journal} {\bibinfo
  {journal} {Progress of Theoretical Physics}\ }\textbf {\bibinfo {volume}
  {127}},\ \bibinfo {pages} {937} (\bibinfo {year} {2012})}\BibitemShut
  {NoStop}%
\bibitem [{\citenamefont {Jackiw}\ and\ \citenamefont
  {Rebbi}(1976)}]{Jackiw-Rebbi}%
  \BibitemOpen
  \bibfield  {author} {\bibinfo {author} {\bibfnamefont {R.}~\bibnamefont
  {Jackiw}}\ and\ \bibinfo {author} {\bibfnamefont {C.}~\bibnamefont {Rebbi}},\
  }\href {https://doi.org/10.1103/PhysRevD.13.3398} {\bibfield  {journal}
  {\bibinfo  {journal} {Phys. Rev. D}\ }\textbf {\bibinfo {volume} {13}},\
  \bibinfo {pages} {3398} (\bibinfo {year} {1976})}\BibitemShut {NoStop}%
\bibitem [{\citenamefont {Shen}(2017)}]{BookShen2017}%
  \BibitemOpen
  \bibfield  {author} {\bibinfo {author} {\bibfnamefont {S.-Q.}\ \bibnamefont
  {Shen}},\ }in\ \href {https://doi.org/10.1007/978-981-10-4606-3_2} {\emph
  {\bibinfo {booktitle} {Topological Insulators: Dirac Equation in Condensed
  Matter}}}\ (\bibinfo  {publisher} {Springer Singapore},\ \bibinfo {address}
  {Singapore},\ \bibinfo {year} {2017})\ pp.\ \bibinfo {pages}
  {17--32}\BibitemShut {NoStop}%
\bibitem [{\citenamefont {Weidemann}\ \emph {et~al.}(2022)\citenamefont
  {Weidemann}, \citenamefont {Kremer}, \citenamefont {Longhi},\ and\
  \citenamefont {Szameit}}]{ExpQC_Weidemann2022}%
  \BibitemOpen
  \bibfield  {author} {\bibinfo {author} {\bibfnamefont {S.}~\bibnamefont
  {Weidemann}}, \bibinfo {author} {\bibfnamefont {M.}~\bibnamefont {Kremer}},
  \bibinfo {author} {\bibfnamefont {S.}~\bibnamefont {Longhi}},\ and\ \bibinfo
  {author} {\bibfnamefont {A.}~\bibnamefont {Szameit}},\ }\href
  {https://doi.org/10.1038/s41586-021-04253-0} {\bibfield  {journal} {\bibinfo
  {journal} {Nature}\ }\textbf {\bibinfo {volume} {601}},\ \bibinfo {pages}
  {354} (\bibinfo {year} {2022})}\BibitemShut {NoStop}%
\bibitem [{\citenamefont {Lin}\ \emph {et~al.}(2022)\citenamefont {Lin},
  \citenamefont {Li}, \citenamefont {Xiao}, \citenamefont {Wang}, \citenamefont
  {Yi},\ and\ \citenamefont {Xue}}]{ExpQC_Quan2022}%
  \BibitemOpen
  \bibfield  {author} {\bibinfo {author} {\bibfnamefont {Q.}~\bibnamefont
  {Lin}}, \bibinfo {author} {\bibfnamefont {T.}~\bibnamefont {Li}}, \bibinfo
  {author} {\bibfnamefont {L.}~\bibnamefont {Xiao}}, \bibinfo {author}
  {\bibfnamefont {K.}~\bibnamefont {Wang}}, \bibinfo {author} {\bibfnamefont
  {W.}~\bibnamefont {Yi}},\ and\ \bibinfo {author} {\bibfnamefont
  {P.}~\bibnamefont {Xue}},\ }\href
  {https://doi.org/10.1103/PhysRevLett.129.113601} {\bibfield  {journal}
  {\bibinfo  {journal} {Phys. Rev. Lett.}\ }\textbf {\bibinfo {volume} {129}},\
  \bibinfo {pages} {113601} (\bibinfo {year} {2022})}\BibitemShut {NoStop}%
\end{thebibliography}%
\appendix
\onecolumngrid
\section{Symmetry Analysis}
In this section, we investigate the symmetry properties of the non-Hermitian BHZ model and the mass term induced SOTI model $\hamiltonian{NH-SOTI}$ defined on a 2D QL with AB tiling. Table \ref{Table_Symm} shows the symmetries of the three  Hamiltonians: $\hamiltonian{NH-BHZ}$ (Eq.~\eqref{eqn:nonhermitian_bhz_hamiltonian}),  $\hamiltonian{NH-SOTI} = \hamiltonian{NH-BHZ} + \hamiltonian{M}$ (Eq.~\eqref{eq:NH-SOTI}), and $\hamiltonian{NH-SOTI}' = \hamiltonian{NH-BHZ} + \hamiltonian{M}'$ (Eq.~\eqref{ruichen_mass_term}). The Hamiltonian $\hamiltonian{NH-BHZ}$ respects a variant of time-reversal symmetry in non-Hermitian systems (TRS$^\dagger$): $\mathcal{T} = U_{\mathcal{T}}T $, a variant of particle-hole symmetry (PHS$^\dagger$): $\mathcal{C} = U_{\mathcal{C}}\mathcal{K}$  and, thus, chiral symmetry: $\mathcal{S} = \mathcal{TC}$. Here, the unitary matrices $U_{\mathcal{T,C}}$ satisfy $U_{\mathcal{T}}U_{\mathcal{T}}^* = -1$,  $U_{\mathcal{C}}U_{\mathcal{C}}^* = 1$, and, $T$ and $\mathcal{K}$ denotes transposition and complex conjugation respectively. $m_x,m_y$ and $m_z$ represents the mirror symmetries reflecting the QL about $x,y$ and $z$ respectively. $\mathcal{P}$ denotes the parity operator (spatial inversion). The Hamiltonian $\hamiltonian{NH-SOTI}$ breaks both $\mathcal{T}$ and $\mathcal{C}$ but preserves the combined symmetry $\mathcal{S}$, whereas  $\hamiltonian{NH-SOTI}'$ preserves $\mathcal{C}$. We found that the zero energy modes (ZEMs) of $\hamiltonian{NH-SOTI}$ are most likely protected by the combined symmetry $m_z\mathcal{C}$ while the ZEMs of $\hamiltonian{NH-SOTI}'$ are protected by the combined symmetry of $\mathcal{S}$ and $\eta$. 

\begin{longtable}{|c|c|c|c|c|}
    \hline
    Symmetry & Condition on $H$ & $\hamiltonian{NH-BHZ}$ & $\hamiltonian{NH-SOTI }$ & $\hamiltonian{NH-SOTI}'$ \\
    \hline
    $\text{TRS}^\dagger = \mathcal{T} = U_{\mathcal{T}}T$  & $U_{\mathcal{T}}H^{T}U_{\mathcal{T}}^{-1} = H$ & \checkmark & $\times$ & $\times$\\
    \hline
    $\text{TRS} = \mathcal{T}' = U_{\mathcal{T}}\mathcal{K}$  & $U_{\mathcal{T}}H^{*}U_{\mathcal{T}}^{-1} = H$ & $\times$ & $\times$ & $\times$\\
    \hline
    $\text{PHS}^\dagger = \mathcal{C} = U_{\mathcal{C}}\mathcal{K}$  & $U_{\mathcal{C}}H^{*}U_{\mathcal{C}}^{-1} = -H$ & \checkmark & $\times$ & \checkmark \\
    \hline
    $\text{PHS} = \mathcal{C}' = U_{\mathcal{C}}T$  & $U_{\mathcal{C}}H^{T}U_{\mathcal{C}}^{-1} = -H$ & $\times$ & $\times$ & $\times$ \\
    \hline
    $\mathcal{S} = \mathcal{T}\mathcal{C}$ & $U_{\mathcal{S}}H^{\dagger}U_{\mathcal{S}}^{-1} = -H$ & \checkmark & \checkmark & $\times$ \\
    \hline
    $\eta = \sigma_3\tau_0$  & $\eta H^{\dagger}\eta^{-1} = H$ & \checkmark & $\times$ & $\times$ \\
    \hline
    $m_x = U_{m_x}\mathcal{M}_x$  & $m_xHm_x^{-1} = H$ & $\times$ & $\times$ & $\times$ \\
    \hline
    $m_y = U_{m_y}\mathcal{M}_y$  & $m_yHm_y^{-1} = H$ & \checkmark & $\times$ & \checkmark \\
    \hline
    $m_z = U_{m_z}$  & $m_zHm_z^{-1} = H$ & $\times$ & $\times$ & $\times$ \\
    \hline
    $m_zm_x = U_{m_z}U_{m_x}\mathcal{M}_x$ & $m_zm_xH{(m_zm_x)}^{-1} = H$ & \checkmark & \checkmark & $\times$ \\
    \hline
    $\mathcal{P} = U_{\mathcal{P}}\mathcal{I}_{xy}$ & $\mathcal{P}H\mathcal{P}^{-1} = H$ & \checkmark & $\times$ & $\times$\\
    \hline
    $m_x\mathcal{T}' = U_{m_x}\mathcal{M}_xU_{\mathcal{T}}\mathcal{K}$  & $m_xU_{\mathcal{T}}H^*(m_xU_{\mathcal{T}})^{-1} = H$ & \checkmark & \checkmark & $\times$ \\
    \hline
    $m_x\mathcal{C}' = U_{m_x}\mathcal{M}_xU_{\mathcal{C}}T$  & $m_xU_{\mathcal{C}}H^T(m_xU_{\mathcal{C}})^{-1} = -H$ & \checkmark & \checkmark & $\times$ \\
    \hline
    $m_x\mathcal{C} = U_{m_x}\mathcal{M}_xU_{\mathcal{C}}\mathcal{K}$  & $m_xU_{\mathcal{C}}H^*(m_xU_{\mathcal{C}})^{-1} = -H$ & \checkmark & $\times$ & \checkmark \\
    \hline
    $m_y\mathcal{T} = U_{m_y}\mathcal{M}_yU_{\mathcal{T}}T$  & $m_yU_{\mathcal{T}}H^T(m_yU_{\mathcal{T}})^{-1} = H$ & \checkmark & \checkmark & $\times$ \\
    \hline
    $m_z\mathcal{C}' = U_{m_z}U_{\mathcal{C}}T$  & $m_zU_{\mathcal{C}}H^T(m_zU_{\mathcal{C}})^{-1} = -H$ & \checkmark & \checkmark & $\times$ \\
    \hline
    $m_z\mathcal{T}' = U_{m_z}U_{\mathcal{T}}\mathcal{K}$  & $m_zU_{\mathcal{T}}H^*(m_zU_{\mathcal{T}})^{-1} = H$ & \checkmark & $\times$ & \checkmark \\
    \hline
    \caption{Symmetries of $\hamiltonian{NH-BHZ}$, $\hamiltonian{NH-SOTI}$ and $\hamiltonian{NH-SOTI}'$ on a square QL. Here the unitary matrices are $U_{\mathcal{T}} = i\sigma_2\tau_0$, $U_{\mathcal{C}} = \sigma_3\tau_1$, $U_{m_x} = \sigma_1\tau_0$, $U_{m_y} = \sigma_2\tau_3$, $U_{m_z} = \sigma_3\tau_0$ and $U_{\mathcal{P}} = \sigma_0\tau_3$. The matrices $\mathcal{M}_x,\mathcal{M}_y$, and $\mathcal{I}_{xy}$ are orthogonal matrices permuting the sites of the QL to flip the lattice vertically, horizontally and both combined, respectively.}
    \label{Table_Symm}
\end{longtable}
\section{Addition of a non-Hermitian on-site gain-and-loss term to Hermitian BHZ model \protect}
    In the main text, we explore the reality of the spectra and the corner states in non-Hermitian SOTI models obtained through the addition of two different mass terms to the non-Hermitian BHZ model. Here, we take an alternative path. We start with a Hermitian SOTI defined on a QL   \cite{RuiChen_etal_prl} and add a non-Hermitian onsite gain-and-loss term. Our goal is twofold: $(1)$ To check if the corner modes are robust to the inclusion of non-Hermiticity. $(2)$ If they turn out to be robust, investigate the reality of the corresponding spectrum. 
    
    The Hamiltonian for Hermitian SOTI can be written as \cite{RuiChen_etal_prl}
    \begin{align}\label{eqn:Hermitian_SOTI}
        \hamiltonian{SOTI} = \sum_{m\neq n} \hat{c}^\dagger_mH'_{mn}\hat{c}_n + \sum_n \hat{c}^\dagger_nH'_n \hat{c}_n,
    \end{align}
    with the hopping term $H'_{mn}$ given by
    \begin{equation}\label{hoppingSOTI}
        H'_{mn} = - \frac{f(r_{mn})}{2}\Bigl[it_1\bigl(\sigma_3\tau_1\text{cos}\phi_{mn} + \sigma_0\tau_2\text{sin}\phi_{mn}\bigr) +t_2\sigma_0\tau_3 - g\sigma_1\tau_1\text{cos}(2\phi_{mn})\Bigr],  
    \end{equation}
    and on-site term
    \begin{equation}\label{onsiteSOTI}
        H'_{n} = (M+2t_2)\sigma_0\tau_3,
    \end{equation}
      with all the parameters retaining their meaning from Sec.~\ref{subsec:nonhermitianbhzmodel}. The Hamiltonian in \eqref{eqn:Hermitian_SOTI} preserves particle-hole symmetry $\mathcal{C}$, defined by $\mathcal{C} =\sigma_3\tau_1\mathcal{K}$, with $\mathcal{K}$ denoting conjugation. The model supports four corner states that are protected by the combined symmetries $\mathcal{C}$ and $C_4m_z$, with $C_4$ denoting fourfold rotation symmetry, and $m_z$ denoting mirror symmetry. We introduce an on-site non-Hermitian term representing gain-and-loss,
     \begin{equation}\label{loss-gain term}
         \hamiltonian{loss-gain} = \sum_n \hat{c}^\dagger_n\bigl(i\gamma\sigma_3\tau_3\bigr)\hat{c}_n.
     \end{equation}
    
\begin{figure}[H]
        \centering
        \begin{minipage}{0.68\columnwidth}
            \centering
            \includegraphics[width=\columnwidth]{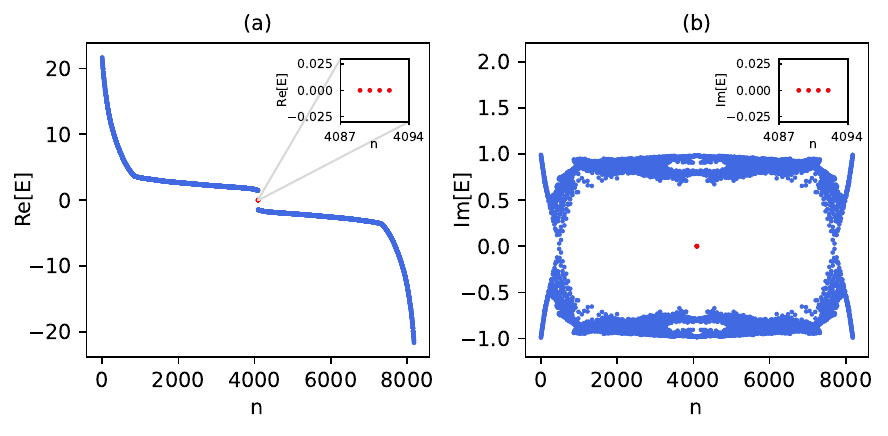} 
            \label{fig:SpectraSupp}
        \end{minipage}\hfill
        \begin{minipage}{0.3\columnwidth}
            \centering
            \includegraphics[width=\columnwidth]{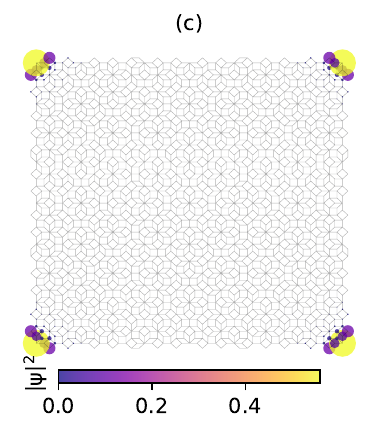} 
            \label{fig:CornerSupp}
        \end{minipage}
        \caption{\label{fig:appendix1} The complex energy spectrum and corner states of $\hamiltonian{SOTI} + \hamiltonian{loss-gain}$ (Eqs.~\eqref{eqn:Hermitian_SOTI} and \eqref{loss-gain term})
        with open boundary conditions. Panels (a) and (b) show the real and imaginary part of the spectrum versus the eigenvalue index $n$ for $\gamma =1$ with insets focussing on ZEMs. (c) displays the wavefunction probability density $\sum_{n\in \text{ZEMs}}|\psi_n|^2$ of the ZEMs localized on the corner of the quasicrystalline lattice. Here $\gamma = 1$ corresponds to the topologically non-trival phase.} 
    \end{figure}

    We diagonalize the Hamiltonian matrix under open boundary conditions(OBC) with $\xi = 1, t_1 = t_2 = 1, g = 1, \text{and } M = -1$. We choose $\gamma = 1$, corresponding to the topological non-trivial phase hosting corner modes. The spectrum and the probability distribution of the zero energy modes (ZEMs) are displayed in Fig. \ref{fig:appendix1}. 
    
    We immediately observe from Fig. \hyperref[fig:appendix1]{9(a,b)} that the spectrum is complex. Also, from Fig. \hyperref[fig:appendix1]{9(c)}, we notice that there is no asymmetry of the corresponding corner modes unlike the models discussed in the main text. The zero energy modes are protected by the particle-hole symmetry $\mathcal{C}'$. 
\end{document}